\documentclass[11pt,preprint]{aastex}

\def\be{\begin{equation}}
\def\en{\end{equation}}

\def\msun{M_{\sun}}

\def\lsun{L_{\sun}}
\def\msunyr{\rm{M_{\sun} \, yr^{-1}}}

\def\mdot{\dot{M}}

\def\h2{\rm{H$_2$}}

\begin{document}

\title
{Evolution of X-ray and FUV Disk-Dispersing Radiation Fields$^*$\footnotetext[1]{This paper includes data gathered with the 6.5 meter Magellan Telescopes located at Las Campanas Observatory, Chile.}}

\author{Laura Ingleby\altaffilmark{1}, Nuria Calvet\altaffilmark{1}, Jesus Hern{\'a}ndez\altaffilmark{2}, Cesar Brice{\~n}o\altaffilmark{2}, Catherine Espaillat\altaffilmark{3}, Jon Miller\altaffilmark{1},  Edwin Bergin\altaffilmark{1}, Lee Hartmann\altaffilmark{1}
}

\altaffiltext{1}{Department of Astronomy, University of Michigan, 830 Dennison Building, 500 Church Street, Ann Arbor, MI 48109, USA; lingleby@umich.edu, ncalvet@umich.edu, jonmm@umich.edu, ebergin@umich.edu, lhartm@umich.edu}
\altaffiltext{2}{Centro de Investigaciones de Astronom{\'i}a (CIDA), M{\'e}rida, 5101-A, Venezuela; jesush@cida.ve, briceno@cida.ve}
\altaffiltext{3}{NSF Astronomy and Astrophysics Postdoctoral Fellow.  Harvard-Smithsonian Center for Astrophysics, 60 Garden Street, MS-78, Cambridge, MA 02138, USA; cespaillat@cfa.harvard.edu}

\begin{abstract}
We present new X-ray and Far Ultraviolet (FUV) observations of T Tauri stars covering the age range 1 to 10 Myr.  Our goals are to observationally constrain the intensity of radiation fields responsible for evaporating gas from the circumstellar disk and to assess the feasibility of current photoevaporation models, focusing on X-ray and UV radiation.  We greatly increase the number of 7--10 Myr old T Tauri stars observed in the X-rays by including observations of the well populated 25 Ori aggregate in the Orion OB1a subassociation.  With these new 7--10 Myr objects, we confirm that X-ray emission remains constant from 1--10 Myr.  We also show, for the first time, observational evidence for the evolution of FUV radiation fields with a sample of 56 accreting and non-accreting young stars spanning 1 Myr to 1 Gyr.  We find that the FUV emission decreases on timescales consistent with the decline of accretion in classical T Tauri stars until reaching the chromospheric level in weak T Tauri stars and debris disks.  Overall, we find that the observed strength of high energy radiation is consistent with that required by photoevaporation models to dissipate the disks in timescales of approximately 10 Myr.  Finally, we find that the high energy fields that affect gas evolution are not similarly affecting dust evolution; in particular, we find that disks with inner clearings, the transitional disks, have similar levels of FUV emission as full disks.

\end{abstract}


\section{ Introduction} 

Circumstellar disks, formed as a consequence of the star formation process, evolve from an initially gas rich phase to a system composed of small, planetary sized bodies, with little remaining gas.  The physical processes responsible for depleting circumstellar gas from the initial minimum mass solar nebula (MMSN) to the low quantities remaining in debris disks, those with significant dust evolution and gas depletion, have been the focus of gas dispersal models \citep{clarke01,alexander06,gorti09,gorti09b,owen10}.  Models for the evolution of the dust \citep{dullemond05,dalessio06,robitaille07} have been successfully utilized in the interpretation of a variety of Infrared (IR) spectral shapes \citep{sargent06,hernandez07,espaillat07,currie08,furlan09,najita10}.  These models indicate that while some of the material is locked into planets and still more is accreted onto the star through viscous evolution, it is likely that photoevaporation is the most significant process playing a role in the dispersal of the gas disk.  There are several contributors to photoevaporation, including X-rays or Extreme Ultraviolet (EUV) photons from the central star and Far Ultraviolet (FUV) emission produced in an accretion shock at the stellar surface of classical T Tauri stars, or CTTS \citep{owen10,gorti09b}.  At any given time, contributions from several different photoevaporation mechanisms will occur, but the overall timescales when each contribution is greatest may vary.  For example, accretion related radiation fields will decay as the mass accretion rate ($\mdot$) decreases.  Alternatively, X-ray emission may increase as the accretion related radiation fields drop due to increased magnetic activity in non-accreting T Tauri stars (WTTS), observed as a possible increase in X-ray luminosity from 1-100 Myr in the \emph{Chandra} Orion Ultradeep Project (COUP) survey of young stars \citep{preibisch05}.  The changing environment may play a role, as FUV and X-rays dominate photoevaporation early at $\mdot<10^{-6}\; \msunyr$, while EUV photons are absorbed in the accretion streams and do not become a significant contributor until $\mdot$ drops to $<10^{-8}\; \msunyr$  \citep{gorti09}. 

With the large number of factors to consider in gas disk evolution, the intensity of UV and X-ray radiation fields and the timescales relevant to their evolution are essential parameters in the theory of photoevaporation.  Not only observations of the radiation fields, but observations of the frequency of circumstellar disks place constraints on disk evolution theory, with only 10\% of sources retaining disks at 10 Myr \citep{hernandez07b}.  Full (or primordial) disks and debris disks are observed at both 5 and 10 Myr \citep{hernandez06,hernandez09}, indicating that disk dispersal mechanisms must occur on a number of timescales in a single star forming region.  Transitional and pre-transitional disks, which due to the presence of a gap are interpreted as an intermediary stage between full and debris disks \citep{strom89,calvet02,espaillat08}, are observed to have a low frequency indicating that when disk dissipation begins it occurs quickly, taking as little as 0.5 Myr \citep{luhman10,hernandez10,muzerolle10}.  Alternatively, disk evolution may proceed along another path in which a similar decline in IR fluxes from all disk radii indicates that no gap or hole is formed \citep{sicilia10,manoj10,hernandez10}.  With the diversity observed in disk evolution, different dispersal mechanisms may be necessary.  In this paper we investigate the relevant timescales for the evolution of disk dispersing emission from T Tauri stars with new X-ray and FUV observations spanning 1-10 Myr, and include previous observations of a larger sample of observations extending to young stars at 1 Gyr.

While EUV emission is almost impossible to observe due to interstellar hydrogen absorption, X-ray observations are numerous.  X-ray emission in T Tauri stars is significantly stronger, up to 1000 times, than the X-ray emission from main sequence stars, and comes from hot plasma with temperatures up to 60 MK or energies up to 5 keV \citep{feigelson02}.  The current paradigm is that this hot plasma is produced through enhanced magnetic activity, with stronger magnetic reconnection events the likely cause of observed short period flaring activity \citep{feigelson93}.  There is some evidence that strong soft X-ray emission is produced in an accretion shock on the stellar surface \citep{gunther07}.  Currently observations of a soft X-ray component which dominates over the coronal emission are unique to the star TW Hya, although other sources have exhibited accretion produced X-ray emission, though at a significantly lower level \citep{robrade06}.  Surveys of young, $<$10 Myr star forming regions have shown no evidence of a decline in the X-ray emission as $\mdot$ decreases, indicating that accretion related X-ray emission is not a significant contributor to the total X-ray flux in most circumstances \citep{preibisch05,mercer09}.  While roughly constant between 1 and 10 Myr, eventually the X-ray emission does decrease.  When this decline begins has not been well characterized so far because the only 10 Myr old X-ray sources studied were the eleven sources in the TW Hydra Association \citep[TWA]{webb99}.  Previously, the cluster nearest in age to TWA with well characterized X-ray properties was the 6--8 Myr $\eta$ Chamaeleon cluster \citep{lopez10}, with fourteen sources.  In this paper we analyze X-ray observations from another older cluster, the 25 Ori aggregate in the Orion OB1a subassociation \citep{briceno07}.  At 7--10 Myr, 25 Ori still has 4-5 remaining CTTS and a large number of WTTS, making it the most populous $\sim$10 Myr old star forming region known.  Extinction in this off-cloud area is low, making it an ideal region for studying X-rays from T Tauri stars in older, evolved populations.

Less is known about FUV emission from T Tauri stars than X-ray emission because, until recent instruments like the Cosmic Origins Spectrograph (COS) on the Hubble Space Telescope (HST), most FUV detectors lacked the sensitivity to observe dim sources.  Large FUV samples have been obtained with the International Ultraviolet Explorer (IUE) through which FUV line emission has been well characterized \citep{valenti00,johnskrull00}, but little to no signal was achieved in the continuum at $\lambda<1750$ {\AA} for the WTTS.  Space Telescope Imaging Spectrograph (STIS) observations of T Tauri stars have also been analyzed; however, long exposure times were required for all but the brightest sources, limiting the ability of STIS to observe a large population \citep{calvet04, herczeg02,herczeg06}.  Even though IUE and STIS observations are limited, the T Tauri properties observed in these high resolution spectra were used to interpret low resolution spectra obtained in a large sample of CTTS, WTTS, and debris disks using the Advanced Camera for Surveys Solar Blind Channel (ACS/SBC) prism on HST.  While low in resolution, with its excellent sensitivity ACS/SBC obtained some of the first observations of WTTS in the FUV with the continuum detected from 1230 -- 1800 {\AA} \citep{ingleby09}.  

The origin of FUV emission in T Tauri stars has contributions from several sources.  In Ingleby et al. (2011), we show that in addition to an accretion shock component \citep{calvet04}, \h2 in the inner region of CTTS disks contributes to the FUV line and continuum emission \citep{abgrall97,ingleby09,bergin04,herczeg04,herczeg02,france10}.  The evolution of FUV emission has not been well studied, as FUV spectra of T Tauri stars greater than 1-3 Myr are rare, and previously included only four T Tauri stars in the TWA region and older debris disks \citep{ingleby09,herczeg04}.

In this paper, we discuss the evolution of high energy radiation fields by analyzing the X-ray and FUV luminosities of a large sample of sources (the largest FUV sample to date) with ages between 1 Myr and 10 Myr, including a small sample of post T Tauri objects with ages up to 1 Gyr.  In $\S$\ref{obs} we discuss the new X-ray, FUV, and optical observations used in our analysis.  In $\S$\ref{mage} we present new X-ray selected T Tauri stars in 25 Ori and Chamaeleon I.  In $\S$\ref{xspec_text} we present results of spectral analysis performed on high count rate X-ray sources and we discuss the calculation of X-ray and FUV luminosities in $\S$\ref{xraylum} and $\S$\ref{fuvlum}.  Finally, in $\S$\ref{implications} we analyze the high energy radiation fields and discuss whether the observed evolution is consistent with current disk evolution theories.

\section{Source Sample and Observations}
\label{obs}

\subsection{X-ray Data}
\label{xray}

\emph {Chandra} observations of four fields in the 7--10 Myr 25 Ori star forming association were obtained in 2008 January with ACIS-I and S chips.  Observations were done in FAINT mode with 10 ks exposures in GO Program 09200907 (PI: Hartmann).  Additionally, observations of two fields in the 2.5 Myr Chamaeleon I star forming region were obtained in 2009 February with ACIS-I and S chips.  Again, the observations were taken in FAINT mode in GO Program 09200909 (PI: Calvet) with 25 ks exposure times.  We analyzed the evt2 files provided from the \emph{Chandra} version 7.6.11.4 processing pipeline.  X-ray detections were identified using the \emph{Chandra} Interactive Analysis of Observations software \citep[CIAO]{fruscione06} script WAVDETECT with the default significance threshold set to $10^{-6}$.  We cross-correlated the X-ray detections with the \cite{luhman04}, \cite{luhman08}, and the CIDA Variability Survey of Orion \citep[CVSO]{briceno05} optical catalogs and identified previously studied T Tauri sources.

\begin{figure}
\plotone{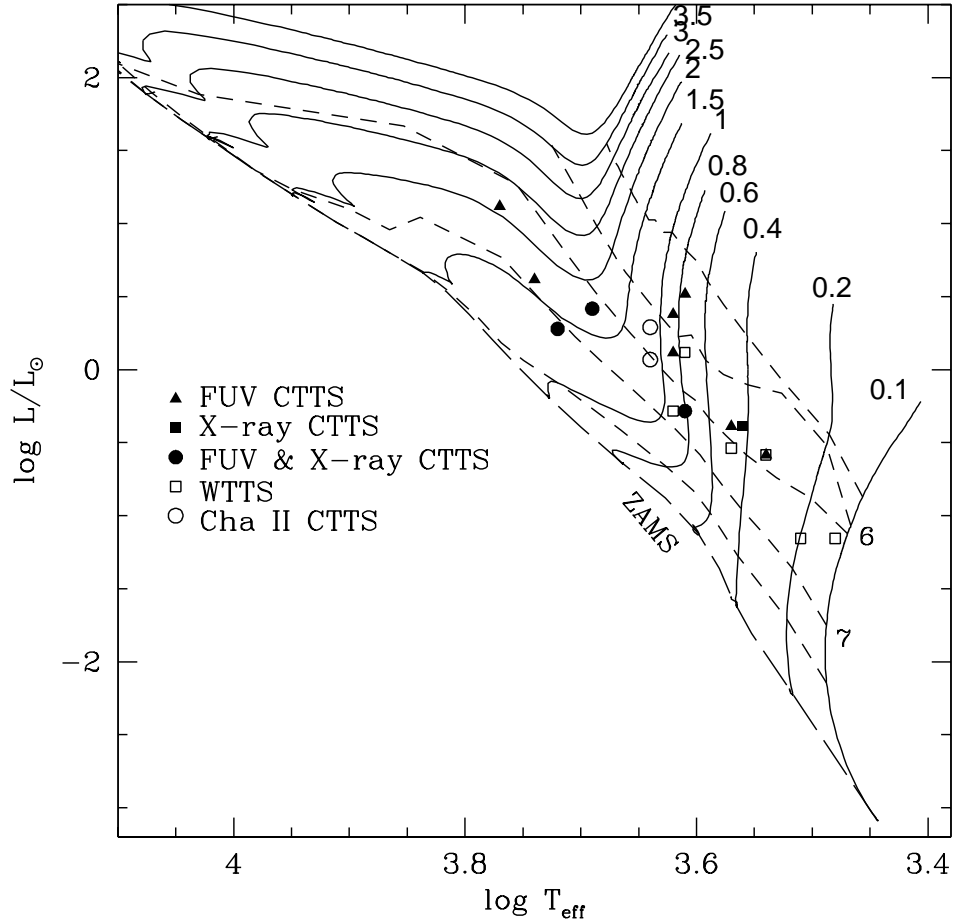}
\caption{H-R diagram of X-ray and FUV samples of Chamaeleon sources.  We show the locations on the H-R diagram of Chamaeleon I X-ray and FUV sources and identify those with both X-ray and FUV observations.  All WTTS were observed in X-rays and do not have FUV spectra.  We also show the location of the older Chamaeleon II sources which were observed with ACS/SBC.  Solid lines are evolutionary tracks from \cite{siess00} and are labeled in units of solar masses.  Dashed lines are isochrones and correspond to log age = 5.5, 6, 6.5, 7, and 7.5 years.}
\label{hrcham}
\end{figure}

In Chamaeleon I, the selected fields focused on two CTTS with disks showing evidence for inner clearing, SZ Cha (T6) and T35 \citep{kim09}.  We detected 15 known T Tauri stars in the two fields (out of $\sim$60 X-ray sources identified by WAVDETECT), five of which are accretors \citep{luhman04}.  The sources are listed in Table \ref{resultstable} and their positions in the H-R diagram are shown in Figure \ref{hrcham} based on spectral types and luminosities from the literature (see references in Table \ref{resultstable}).  Using the evolutionary tracks of \cite{siess00}, we determine that the sample covers a large range of T Tauri masses from a 1--1.5 $\msun$ upper limit down to almost substellar masses, with a lower limit of 0.1--0.2 $\msun$.  This mass range is similar to that used in \cite{preibisch05}, in which the evolution of X-ray emission in the low mass population of the COUP sources was analyzed.

\begin{figure}
\plotone{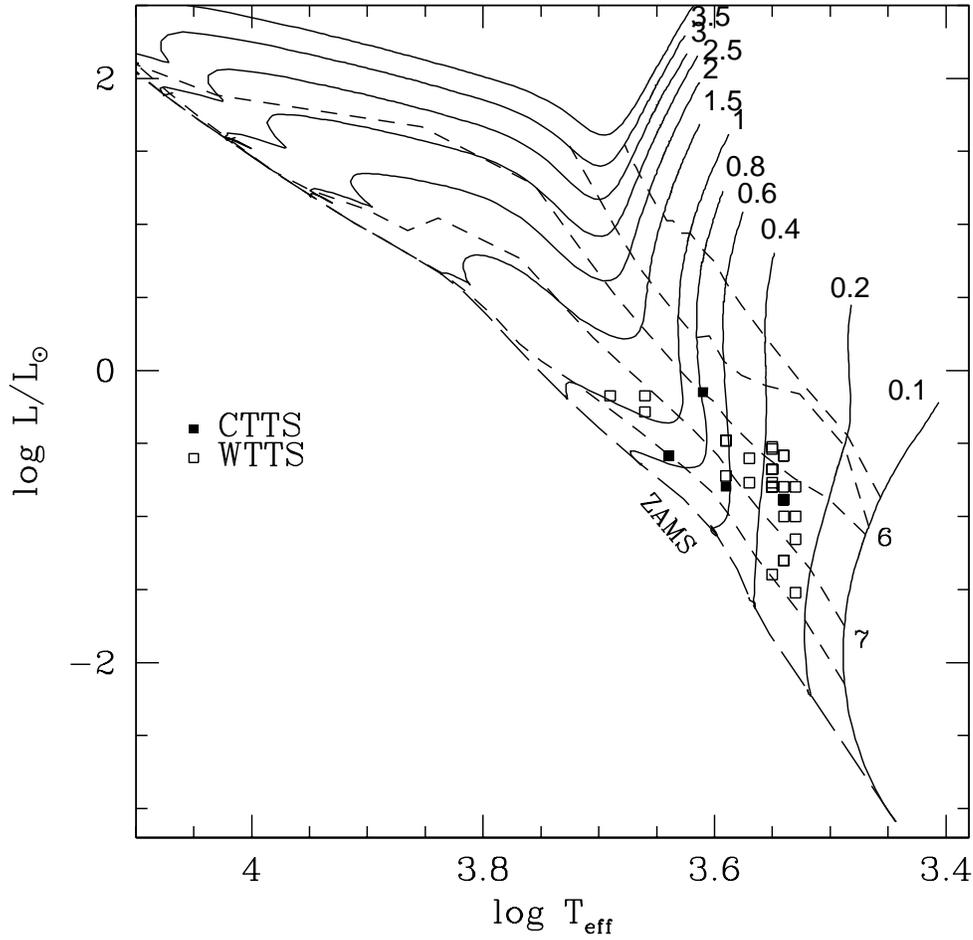}
\caption{H-R diagram of 25 Ori association X-ray detections.  We show the location of all 25 Ori X-ray sources on the H-R diagram with evolutionary tracks and isochrones as defined in Figure \ref{hrcham}.  Sources fall primarily between the 6.5 and 7 Myr isochrones with earlier (hotter) sources appearing older due to birth line effects \citep{hartmann03}.}
\label{hrdiagram}
\end{figure}

Each of the four fields in the 25 Ori aggregate was centered on a known CTTS, specifically CVSO 35, CVSO 206, OB1a 1630 \citep{briceno05,hernandez07}, and the transitional disk CVSO 224 \citep{espaillat08}.  Of $\sim$150 X-ray sources located by WAVDETECT, we identified 36 previously known T Tauri stars, in addition to the 4 targeted sources, from the catalogs of \cite{briceno05,briceno07}, Brice{\~n}o et al. (2010), and \cite{hernandez07} for a total of 40 7--10 Myr classical and weak T Tauri stars.  Of the remaining unidentified X-ray detections, six are possible newly detected T Tauri stars (see \S\ref{opt} for further discussion).  The locations of the 25 Ori sources in the H-R diagram are shown in Figure \ref{hrdiagram}.  Sources published in \cite{calvet05} had known spectral types and luminosities but sources published in \cite{briceno07}, \cite{hernandez07}, and Brice{\~n}o et al. (2010) had only spectral types and optical colors.  To get luminosities for those sources, we first used $V-I$ colors to get the reddening ($A_V$) assuming photospheric colors from \cite{kenyon95}.  With $A_V$, we corrected Two Micron All Sky Survey \citep[2MASS]{skrutskie06} $J$ magnitudes for reddening using the Mathis extinction law with $R_V=3.1$ \citep{mathis90} and calculated the bolometric fluxes \citep{kenyon95}.  Our sample of 25 Ori sources is similar to the CTTS masses of Taurus, from $\sim$0.2--1.0 $\msun$.  Table \ref{resultstable} lists the complete set of sources detected in the Chamaeleon I and 25 Ori fields which are discussed in this paper, along with published spectral types and disk classifications.  

After matching X-ray detections to previously identified T Tauri stars, we used the CIAO script DMEXTRACT to obtain the counts within the extraction regions determined by WAVDETECT for each known T Tauri source.  We included a background region defined as a 5'' annulus around each extraction region, allowing us to determine the net counts and calculate the poisson errors which we used to estimate uncertainties in the X-ray luminosities ($\S$\ref{xraylum}).  Of the 15 Chamaeleon sources, 5 had greater than 250 counts and the spectra of these sources were extracted using the CIAO script PSEXTRACT, using the 3$\sigma$ detection regions determined in WAVDETECT, and the 5'' background annuli.  Each extracted spectrum was re-binned to 20 counts per bin using the GRPPHA task in FTOOLS.  The re-binned spectra were then analyzed in the X-ray spectral fitting package, XSPEC \citep{arnaud96}.  Only three of the 25 Ori sources had counts in excess of 250, likely due to the shorter exposure times and greater distance to the sources, and were analyzed with PSEXTRACT and XSPEC.    

\subsection{UV Data}
\label{uvdata}
We present new FUV observations of ten sources in the 2.5 Myr Chamaeleon I and two in the 4 Myr Chamaeleon II star forming regions obtained with the PR130L grating on the Advanced Camera for Surveys (ACS) Solar Blind Channel (SBC).  These observations were completed between 2008 December and 2009 February in GO Program 11145 (PI: Calvet).  The FUV coverage extends from 1230--1800 {\AA} with R$\sim$300 at 1230 {\AA} and $\sim$50 at 1700 {\AA}.  The \emph{Chandra} and ACS/SBC Chamaeleon I samples have some overlap but there are sources unique to each sample.  We show all the Chamaeleon I and II sources in Figure \ref{hrcham}, identifying the X-ray and FUV sources and those included in both samples.  With ACS/SBC we observed sources from  0.3--2.0 $\msun$ in Chamaeleon I and approximately solar mass sources in Chamaeleon II.

We combine these new observations with previously observed ACS/SBC and STIS samples of T Tauri stars analyzed in \cite{ingleby09}, \cite{calvet04}, and \cite{herczeg02,herczeg04}.  The total sample of sources with FUV spectra consists of 56 sources, eleven of which are WTTS or debris disks, and these are listed in Table \ref{resultstable}, with sources observed by STIS distinguished in bold font.

\subsection{Optical Data}
\label{opt}
A large number of X-ray detections corresponded to unidentified counterparts in IR and optical catalogs.  Using 2MASS and CVSO photometry \citep{briceno05}, we produced color-magnitude diagrams in $V$ and $V-J$ to select sources which fell above the main sequence.  In Chamaeleon I, one such object met the color requirements (K. Luhman, personal communication) and in 25 Ori there were six possible new T Tauri stars detected.

For three of the six T Tauri candidates in 25 Ori, we followed up the X-ray observations by obtaining optical spectra with the Magellan Echellete (MagE) on the Magellan- Clay telescope at Las Campanas Observatory \citep{marshall08}.  MagE provides R$\sim$4100 spectra from 3200--9800 {\AA}, including the Li 6707 line, for confirmation that the sources are young, and H$\alpha$ necessary to classify them as T Tauri stars.  In addition to the MagE spectra of new T Tauri stars, we obtained R$\sim$ 35,000 MIKE (The Magellan Inamori Kyocera Echelle) observations of 16 sources in our sample, primarily of accretors in the 25 Ori and Chamaeleon I and II regions.  

The optical data were reduced using the Image Reduction and Analysis Facility \citep[IRAF]{tody93} tasks ``ccdproc", ``apflatten" and ``doecslit".  Bias corrections were completed with ``ccdproc".  Flats were created using ``apflatten", in which the orders were traced using observations of a quartz lamp for the MIKE flat frames and the 10 reddest orders of MagE.  For the five bluest orders of MagE, we observed a Xe lamp with a wide slit (5'' as opposed to the science slit at 1'') and out of focus for the trace.  The task ``doecslit" was used to extract the science spectrum, extinction correct for airmass, subtract the background, and dispersion correct the spectra using a ThAr comparison lamp.  The blaze function was fit using a bright O or B standard star and used to correct for the blaze in the science exposures.

\section{Analysis of Observations}
\label{analysis}

\subsection{Results from Optical Observations}
\label{mage}
\begin{figure}
\plotone{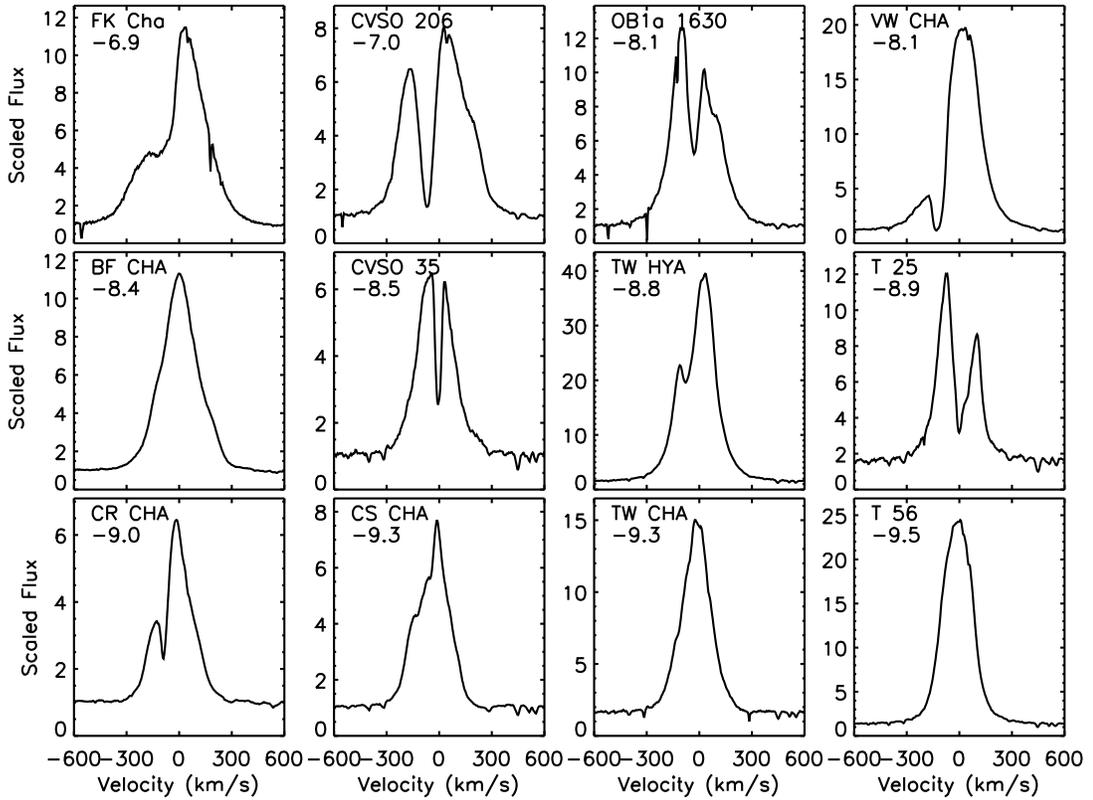}
\caption{H$\alpha$ line profiles from MIKE.  We show the accreting sources in our sample which have MIKE observations, ordered by log $\mdot$ (listed below each source name) calculated from the 10\% line width relation of \cite{natta04}.  The 7--10 Myr sources, CVSO 206, OB1a 1630, and CVSO 35, have remarkably wide and asymmetric H$\alpha$ profiles, indicating that accretion is strong and gas remains in the inner circumstellar disk \citep{ingleby09}.}
\label{halphs}
\end{figure}

MIKE optical spectra were used to confirm status of WTTS or CTTS and obtain accretion rates for the CTTS using the relation between $\mdot$ and H$\alpha$ 10\% width of \cite{natta04}.  We show H$\alpha$ line profiles in Figure \ref{halphs} for the accreting sample of MIKE sources.  The relation between H$\alpha$ width and $\mdot$ is more appropriate for low values of the mass accretion rate; more precise values of the accretion rate will be obtained in a forthcoming analysis of the optical veiling of these sources (Ingleby et al. 2011).

The three candidate members of 25 Ori with MagE optical spectra, discussed in $\S$\ref{opt}, showed Li absorption and but only two (2MASS J05243490+0154207 and 2MASS J05270173+0139157) had weak H$\alpha$ emission, indicating that they are newly identified WTTS.  The third source (2MASS J05244265+0154116) has H$\alpha$ in absorption but may be a young member of 25 Ori still, due to its early spectral type; additional analysis is needed for confirmation.  This final source is not included in any further analysis of the 25 Ori X-ray observations.  The Li and H$\alpha$ equivalent widths, along with the derived spectral types are listed in Table \ref{tabxraynew} for those sources which have optical spectra.  All of the candidate X-ray selected T Tauri stars, in both Chamaeleon I and 25 Ori, are listed in Table \ref{tabxraynew}.  MagE spectra will be obtained for the remaining sources in need of optical spectra.

\subsection{X-ray Spectral Analysis}
\label{xspec_text}
Eight sources in our Chamaeleon I and 25 Ori samples had strong X-ray fluxes allowing for spectral analysis to determine temperatures for the emitting plasmas.  Three of these sources are accreting (CR Cha, SZ Cha, and TW Cha) and the remaining five sources are not \citep{luhman04, calvet05}.  With this sample of high count sources, we looked for any observable distinction between the CTTS and WTTS.  In XSPEC, each source spectrum was fit using either one or two APEC (Astrophysical Plasma Emission Code) models characterized by different temperatures; the choice of models was motivated  by the two-temperature plasma observed in X-ray emission from Pleiades sources \citep{briggs03}.  Assuming a model which includes multiple temperatures also allows for the distinction of a soft accretion related emission component, if present, similar to that observed in TW Hya \citep{gunther07}.  We included an absorption component using the XSPEC model, PHABS, with the equivalent neutral hydrogen column density\footnote{The equivalent neutral hydrogen column density (defined as $N_H$) is the column density of hydrogen required to give the same opacity as that resulting from the true composition of the absorbing gas, which has contributions from He, C, N, O, and Ne in addition to less abundant gasses.} obtained from the HEASARC (High Energy Astrophysics Science Archive Research Center) $N_H$ calculator \citep{kalberla05,dickey90}.  

\begin{figure}
\plotone{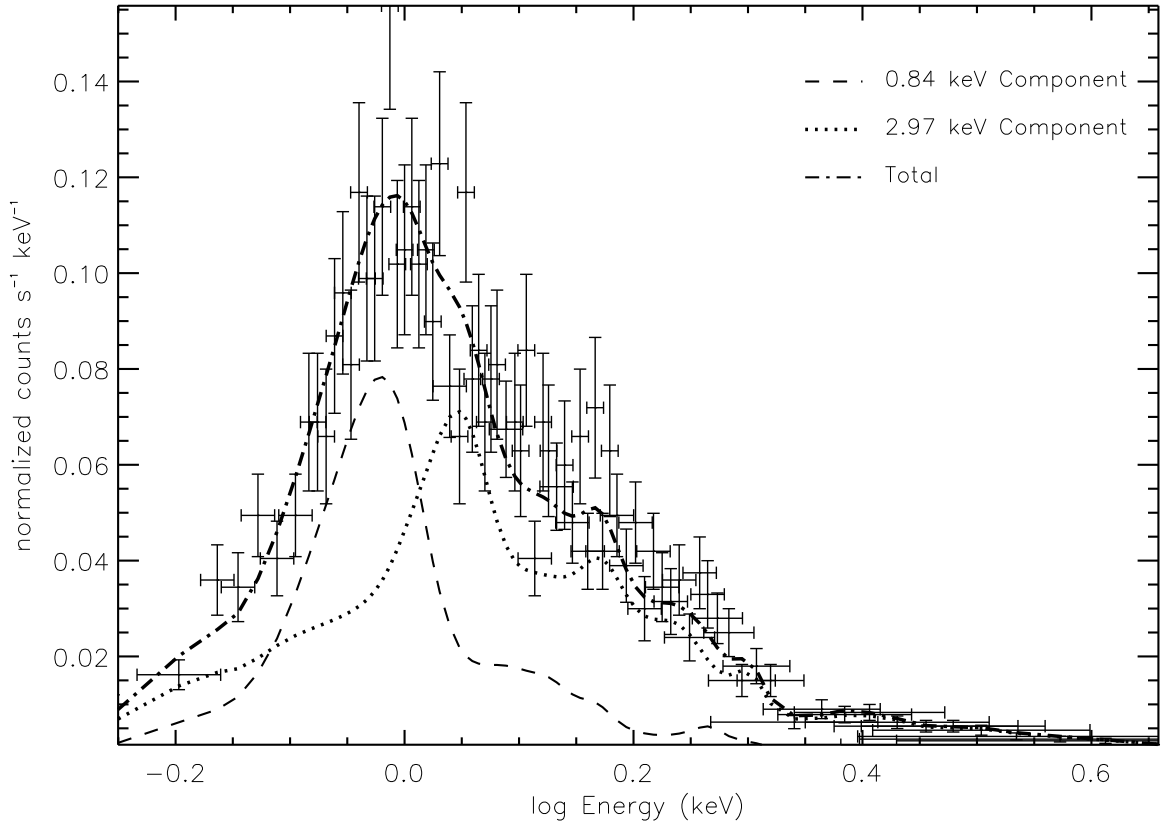}
\caption{\emph{Chandra} ACIS spectrum of the accretor CR Cha.  We show the extracted spectrum of CR Cha (the source with the highest counts in our sample) along with our best APEC model fit to the spectrum.    }
\label{crcha}
\end{figure}

The models with the best statistical fit to the observations (or lowest $\chi^2_{red}$) are given in Table \ref{xspec}.  The best fit models for CR Cha, SY Cha, SZ Cha, and TW Cha produced good $\chi^2_{red}$'s, but the remaining sources had either poor fits or model properties discrepant from the larger sample.  For CHXR 30A, when the absorption model was forced to have a column density of $N_H=8\times10^{20}\;\rm{cm}^{-2}$ \citep{kalberla05,dickey90}, a 64 keV plasma was necessary to fit the spectrum, greater than ten times higher than the highest expected plasma temperatures \citep{feigelson02}.  Allowing the column density to vary, we found a best fit model with $N_H=12.2\times 10^{21}\;\rm{cm}^{-2}$, fifteen times higher than $N_H$ in the four Chamaeleon sources with good fits.  While high, a greater value for $N_H$ is consistent with CHXR 30A being extincted by $A_J\sim3$ magnitudes while the other sources have $A_J=0.0-0.5$ \citep{luhman07}.  The APEC model fits to CVSO 38 resulted in a large temperature for the hot component, a temperature typical of flares in T Tauri stars \citep{getman08}, indicating that the observations may have caught it while flaring.  The APEC components and best model fit to the highest count rate X-ray source (CR Cha) are shown in Figure \ref{crcha}.  For CR Cha we find that a two-temperature model improves the $\chi^2_{red}$ by a factor of 3 compared to a single temperature model.  Like CR Cha, the models of the four sources with the best $\chi^2_{red}$ fits are each comprised of a cool, 0.3-0.9 keV, component and a hotter 2-5 keV component.  We note that the hottest emission (excluding the possible flare component from CVSO 38) comes from SZ Cha which, as a G2, is the source with the earliest spectral type.  

Previous analyses of X-ray spectra from CTTS have shown that the presence of accretion alters the X-ray spectrum by producing emission lines formed in the high densities of the accretion shock \citep{robrade06,brickhouse10}.  Also, cool X-ray components with $kT\sim$0.2--0.3 have been attributed to accretion \citep{gunther07,robrade06}.  We see no clear evidence of accretion related soft X-ray emission in CR Cha, SZ Cha, and TW Cha, the known accretors in our sample.  The cool APEC components are $\sim$2--10 times hotter than the temperatures expected of plasmas heated in an accretion shock, even when considering material falling faster than the free fall velocity which is necessary to produce the soft X-ray shock emission in TW Hya \citep{gunther07}.  \cite{brickhouse10} obtained high resolution Chandra spectra of TW Hya and identified line emission which is attributed to accretion.  Additional observations obtained with XMM-Newton of a sample of four CTTS, including both TW Hya and CR Cha (which we analyze in Figure \ref{crcha}), showed that shock formed line emission may be needed to confirm accretion for sources unlike TW Hya, where the soft accretion component does not strongly exceed the coronal emission.  The resolution of our ACIS spectra prohibited accretion line analysis, which would be informative, particularly for TW Cha, the accretor with a cool component most similar to that in TW Hya.  Interestingly, the sources with the coolest components (SY Cha, 2MASS J05244498+0159465 and RX J0526+0143) are WTTS and have been identified as such by observations of H$\alpha$,  indicating that a detection of a cool component alone cannot confirm accretion and likely line detections are needed as well \citep{robrade06}.

Instead, our results are consistent with the two-temperature coronal emission observed in Pleiades sources \citep{briggs03}; however, for Chamaeleon, the temperatures of the hot component are approximately 2 times higher than the hot components in the Pleiades sources.  This may be indicative of theorized cooling coronal gas, producing a softer spectrum as X-ray sources evolve \citep{kastner03}.  The results of our spectral analysis show that we can use the same treatment to find X-ray luminosities for both the WTTS and CTTS since we see no evidence for an accretion related component.
 
 \subsection{X-ray Luminosities}
 \label{xraylum}

\begin{figure}
\plotone{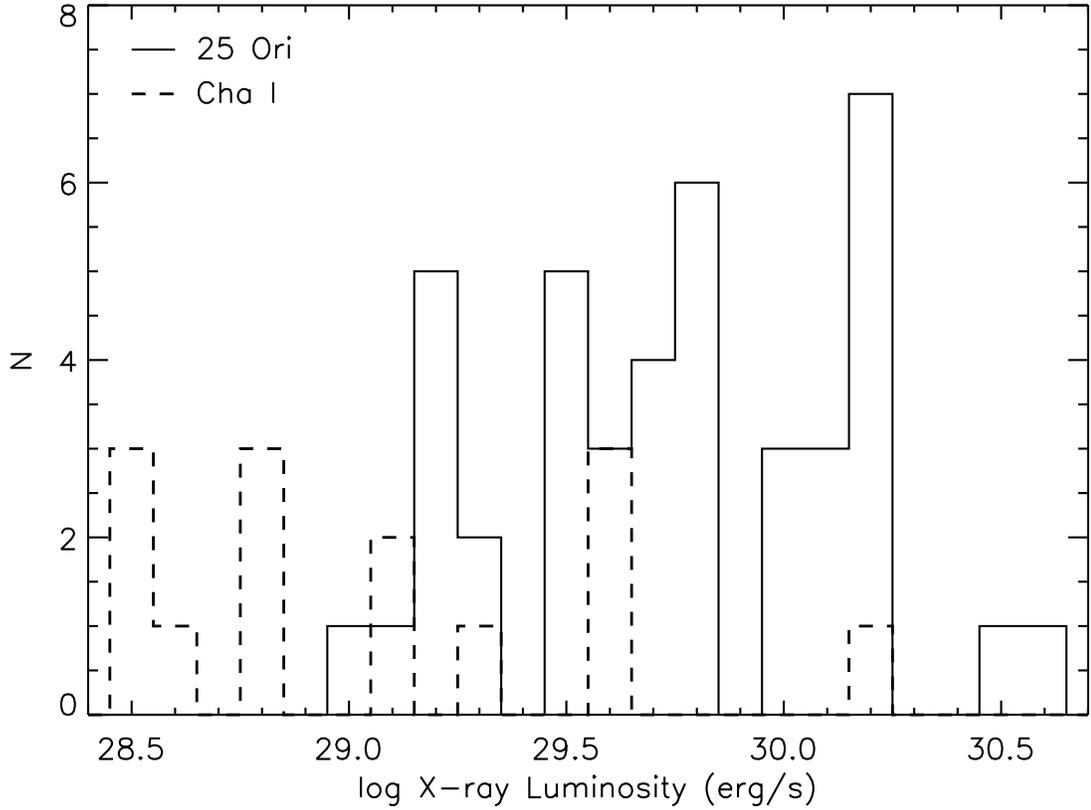}
\caption{Distributions of X-ray luminosities for 25 Ori and Chamaeleon I samples.  The distribution of  X-ray luminosities, corrected for absorption, in each sample does not reproduce the lognormal shape of the X-ray luminosity function found for COUP sources in \cite{feigelson05}.  This indicates that our samples are likely incomplete in mass.  For this reason we supplement our Chamaeleon I observations with previous X-ray observations; however, no additional X-ray observations exist for 25 Ori.}
\label{logNlogS}
\end{figure}

The majority of the X-ray sources had $<$250 counts, so instead of attempting detailed spectral analysis on dim sources we assume typical X-ray temperatures as determined for the sources discussed in $\S$\ref{xspec_text}.  We assume there is no discernible difference between the X-ray properties of the CTTS and WTTS in the spectral region observed by ACIS, as indicated by the results shown in Table \ref{xspec}.  Using WebPIMMS on the HEASARC website we calculated unabsorbed fluxes between 0.2 and 10 keV for the entire sample, correcting for $N_H=0.8\times10^{21}$ and  $N_H=1.1\times10^{21}\;\rm{cm}^{-2}$ for Chamaeleon I and 25 Ori, respectively \citep{kalberla05,dickey90}.  We assumed a single temperature APEC model with kT=2.2 keV and solar abundance.  From the unabsorbed fluxes, X-ray luminosities were calculated assuming distances of 160 and 330 pc for Chamaeleon I and 25 Ori, respectively; the results are listed in Table \ref{resultstable}.  For the sources in Table \ref{xspec}, which were fit with XSPEC, we find that the when using WebPIMMS to find fluxes, the discrepancy is a factor of 2--3.  The distribution of unabsorbed X-ray luminosities is shown in Figure \ref{logNlogS}, which indicates that our Chamaeleon I sample is not complete.  To increase the sample, we include X-ray observations of this region from \cite{feigelson93} in our analysis.  The ROSAT count rates from \cite{feigelson93} were converted to ACIS-S count rates using WebPIMMS before luminosities were calculated.

In addition to the X-ray luminosities calculated here, we include X-ray luminosities from the literature for sources with FUV spectra which were not observed with \emph{Chandra}, primarily in Taurus, TWA, and nearby debris disks (shown in italics in Table \ref{resultstable}).  X-ray fluxes of Taurus sources were calculated as part of the XMM Newton Extended Survey of Taurus  (XEST) project \citep{gudel07}.  \cite{gudel07} gives count rates for the PN camera with the medium filter and these were converted into \emph{Chandra} ACIS-S count rates using WebPIMMS, assuming an input XMM-Newton energy range of 0.4-10 keV and output \emph{Chandra} energy range of 0.2-10 keV, and using the same model parameters as were used to calculate the fluxes from our \emph{Chandra} observations.  X-ray fluxes for sources in Taurus not provided by \cite{gudel07}, along with sources in Chamaeleon I and II, TWA, and field debris disks were converted from ROSAT PSPC count rates \citep{feigelson93,voges99,voges00} into \emph{Chandra} ACIS-S count rates, again using WebPIMMS but using the softer input energy of ROSAT (0.1-2 keV).  Once converted to ACIS-S count rates, the unabsorbed X-ray fluxes were calculated as discussed for our new \emph{Chandra} observations and the luminosities are given in Table \ref{resultstable}.  We also include X-ray luminosities from other young clusters in our analysis of X-ray evolution; those from the $\sim$ 4 Myr cluster, Trumpler 37 \citep{mercer09}, and the 6--8 Myr $\eta$ Cha region \citep{lopez10}.

\subsection{FUV Luminosities}
\label{fuvlum}
Our FUV sample contains sources in Taurus, the Chamaeleon I and II clouds, 25 Ori, TWA, objects in the outskirts of the Orion Nebular Cluster (ONC), and nearby WTTS or debris disks.  We calculated FUV luminosities for these sources between 1230 and 1800 {\AA} after degrading the resolution of the STIS spectra to that of ACS/SBC for comparison.  Spectra were corrected for reddening using $A_V$'s provided in \cite{ingleby09} and \cite{calvet04} for Taurus, ONC, 25 Ori, and the nearby WTTS and debris disks.  For the Chamaeleon I and II sources, $A_V$'s were calculated from published photometry \citep{gauvin92}.  For sources in Taurus, the ONC, and Chamaeleon I and II we corrected for reddening using the extinction curve observed along the line of sight towards HD 29647 with $R_V=3.6$, appropriate for regions with associated molecular clouds \citep{whittet04}.  For sources in the dispersed 25 Ori association, we used the Mathis extinction law with $R_V=3.1$ \citep{mathis90}.  We note that UV extinction laws, especially in the FUV, are very uncertain and introduce the largest error in our FUV luminosity calculations.  We assume a $\pm$0.2 magnitude error in $A_V$, which dominates other sources of error.  For four Chamaeleon sources, we failed to detect the FUV emission and upper limits are provided for these sources.  We calculated the flux in the detector noise and de-reddened the noise by the calculated $A_V$'s.  In this manner, we are accounting for any FUV emission washed out by extinction.  Calculated FUV luminosities are given in Table \ref{resultstable}, with upper limits on the luminosities provided when appropriate.

\section{Implications of Observations}
\label{implications}
\subsection{Evolution of X-ray and UV Emission}
\label{xray_uv}

\begin{figure}
\plotone{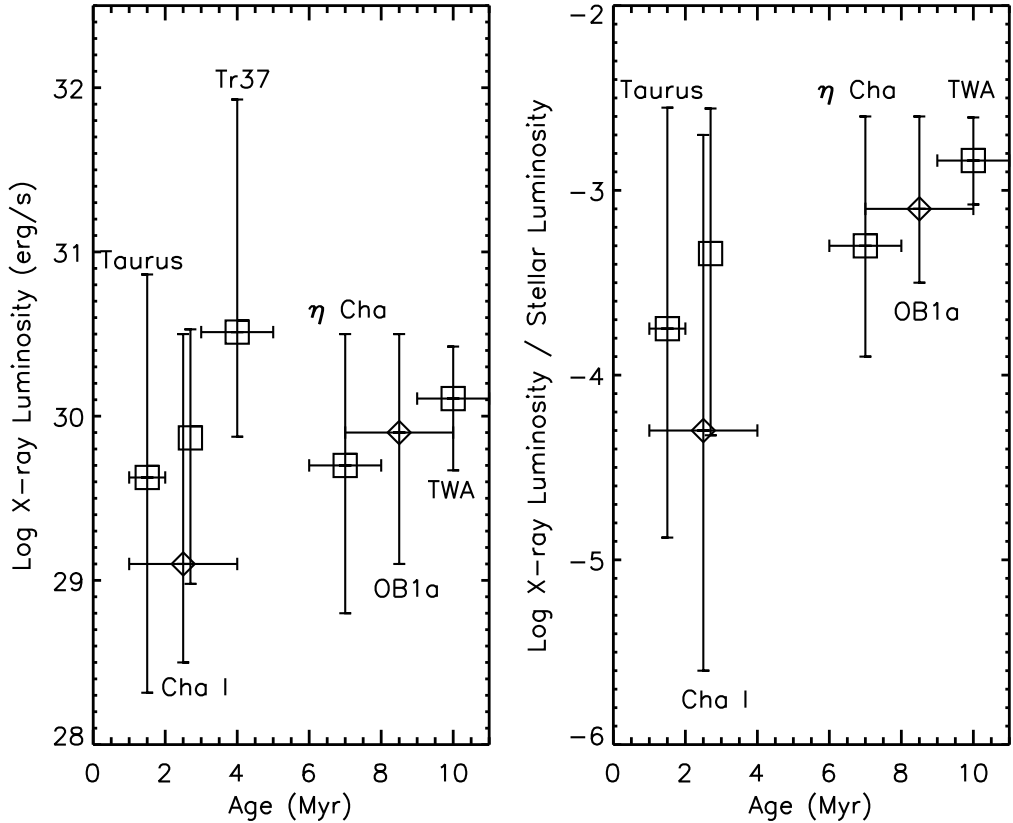}
\caption{\emph{Left:} Median X-ray luminosities of young star forming associations vs age.  For each association, we plot the median and the full range in X-ray luminosity for sources with spectral types K5 and later.  The diamonds represent X-ray luminosities calculated in this work and the boxes represent those taken from the literature: Taurus \citep{gudel07}, Cha I \citep{feigelson93}, Trumpler 37 \citep{mercer09}, $\eta$ Cha \citep{lopez10}, TWA \citep{voges99,voges00}.  We include the sample of \cite{feigelson93} for Cha I (offset in age for clarity) because with only two fields observed here, the Cha I sample was limited.  Our Cha I observations reproduce the range observed by ROSAT well and actually probe dimmer X-ray sources.  \emph{Right:} Normalized X-ray luminosity vs age.  Here we divide the observed X-ray luminosities by the stellar luminosities to normalize by spectral type.  Both figures show that between 1 and 10 Myr, there is no decline in X-ray luminosity.}
\label{xrayvsage}
\end{figure}

The intensity of high energy emission from T Tauri stars has important implications for the dispersion of circumstellar disks and the planets which may be forming within them.  Over a time period of 1-10 Myr, during which these processes are expected to occur, we investigate the evolution of the X-ray and FUV radiation fields.  Figure \ref{xrayvsage} shows the median and full range of X-ray luminosities for the sources in each star forming region with spectral type K5 or later.  The sample of Tr37 may include sources with earlier spectral types (early K); however, spectral types are not well established for this sample.  In addition, it was confirmed that several of the Tr37 sources were flaring during the observations  \citep{mercer09}, contributing to the high X-ray luminosities.  Figure \ref{xrayvsage} reveals that there is no decrease with age in either total X-ray luminosity or the X-ray luminosity normalized by the stellar luminosity, up to 10 Myr.  Most importantly, our addition of the 7--10 Myr 25 Ori X-ray observations supports previous claims that X-ray radiation from the central star remains strong throughout the disk dispersal phase \citep{mercer09,preibisch05}.  While there does appear to be a slight increase in the normalized X-ray luminosity with age, the trend is statistically insignificant, with a Pearson correlation coefficient of only 0.2.  

\begin{figure}
\plotone{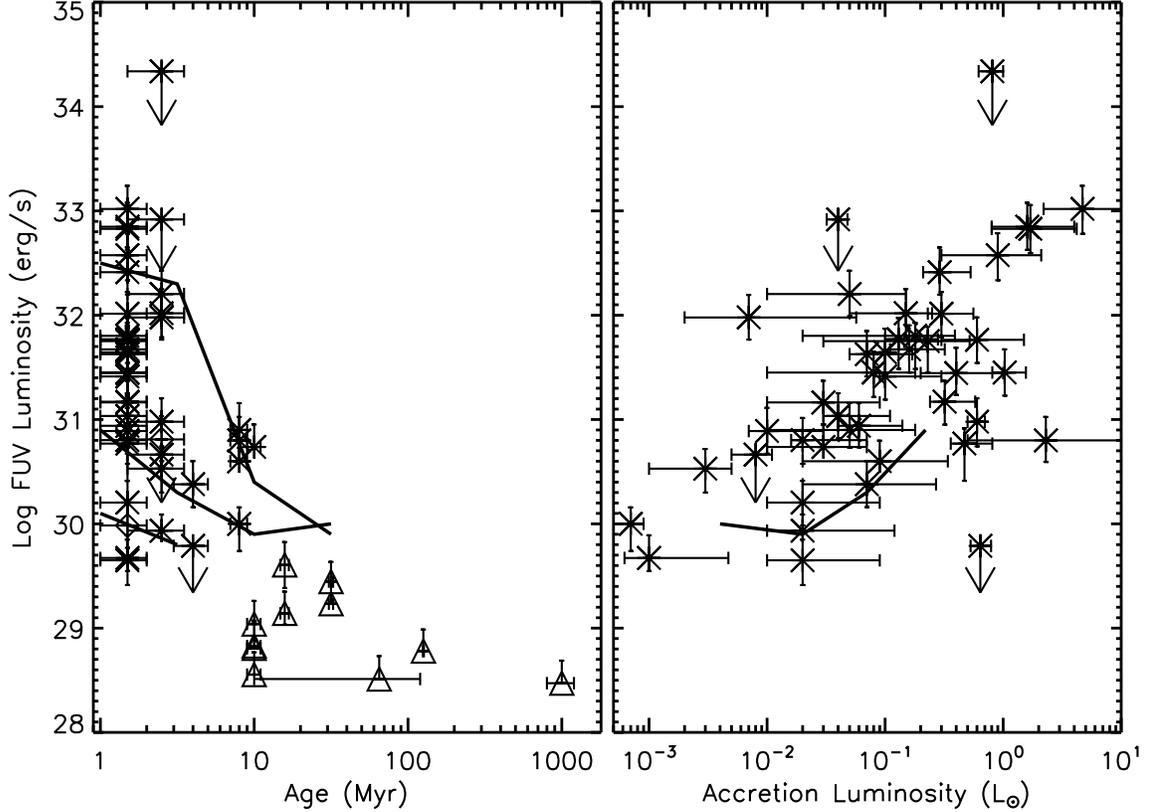}
\caption{\emph{Left:} FUV luminosity vs age for young low-mass stars from the T Tauri phase up to 1 Gyr.  The FUV luminosity of CTTS, displayed as asterisks when accretion rates are known and diamonds when they are not, is observed to decrease with age until becoming entirely chromospheric in WTTS (triangles).  Upper limits were calculated for non-detections, primarily sources with high $A_V$, and are represented by downward arrows.  Solid lines show the predicted FUV luminosity produced in an accretion shock onto the stellar surface \citep{calvet98} as the source evolves and $\mdot$ decreases.  The top and bottom lines are calculated for the observed range in $\mdot$ at each age and the middle line is calculated from $\mdot$ predictions in which a 0.8 $\msun$ CTTS with an initial disk mass of 0.1 $\msun$ evolves viscously \citep{hartmann98,calvet05}.  \emph{Right:} FUV luminosity vs $L_{acc}$.  The solid line shows the predictions of FUV luminosity from accretion onto the 0.8 $\msun$ CTTS through the viscously evolving disk.  Here, we omit the predicted FUV luminosities for the upper and lower limits of observed $\mdot$'s, which were shown on the left.}
\label{fuvvslaccage}
\end{figure}

Figure \ref{fuvvslaccage} shows the evolution of FUV luminosities from 1 Myr T Tauri stars to 1 Gyr young stars and debris disks.  While there is an observed decrease in FUV luminosity out to 1 Gyr, a decrease with age can also be seen among the 1-10 Myr sources, with the WTTS characterized by the lowest FUV luminosities.  By the WTTS phase, FUV luminosities are expected to be entirely chromospheric.  With no additional accretion, FUV emission is no longer produced in an accretion shock and because the end of accretion indicates the depletion of inner disk gas, FUV emission from hot molecular gas no longer contributes to the FUV flux \citep{ingleby09,pascucci06}.  In addition to the decline with age, a correlation between FUV luminosity and $L_{acc}$ is observed, with a Pearson correlation coefficient of 0.78 (calculated with the exclusion of sources with only upper limits), indicating that FUV emission is strongly correlated with $L_{acc}$.  Values of $L_{acc}$ were obtained from the literature (see references in Table \ref{resultstable}) or calculated from H$\alpha$ 10\% line widths \citep[$\S$\ref{mage}]{natta04}.  

This correlation between FUV luminosity and $L_{acc}$ is not surprising, given results that FUV emission is produced in the hot gas of the accretion shock \citep{calvet04,calvet98,herczeg02}.  Therefore, the decrease in FUV luminosity between 1 and 10 Myr can be understood in terms of the decrease of $\mdot$ in this time period \citep{hartmann98,calvet05,sicilia06,fedele10}.  We use the model of \cite{calvet98} to predict the amount of FUV emission produced in the accretion column.  This model assumes that the circumstellar disk is truncated at a few stellar radii by the stellar magnetic field.  At the truncation radius, gas is channeled along the magnetic fields lines from the disk to the star, impacting the stellar surface at approximately the free fall velocity.  Such fast moving material produces an accretion shock on the surface of the star which heats the gas to $\sim$1 MK.  Emission from this hot gas heats both the photosphere below the shock and the pre-shock material, producing an accretion column spectrum with a significant excess over the photospheric emission in the UV \citep{calvet98}.

As inputs to the accretion column model, we assume a 0.8$\msun$ source evolving along the \cite{siess00} evolutionary tracks, from which we obtain the radius, effective temperature, and luminosity of the source from 1--10 Myr.  We assume the percentage of the stellar surface covered by accretion columns, the filling factor, remains constant at 1\% and vary the energy flux in the accretion column for a given $\mdot$ \citep{calvet98}.  Using these quantities we calculate the column emission and integrate the flux from 1230--1800 {\AA} for the highest and lowest $\mdot$'s observed at each age \citep{calvet05}.  We also calculated the FUV accretion emission for the $\mdot$'s predicted from a viscously evolving disk with an initial mass of 0.1$\msun$ \citep{hartmann98}.  Using this technique, we find we can approximate the range of observed FUV luminosities well (left panel of Figure \ref{fuvvslaccage}), but for typical CTTS masses, the predicted FUV fluxes from the viscously evolving disk are too low (right panel of Figure \ref{fuvvslaccage}).  

The missing FUV emission is probably due to several sources which are not included in our accretion column models.  First, emission from \h2 gas was observed in the FUV spectra of CTTS and was necessary to explain the FUV continuum between 1500 and 1700 {\AA}.  The contribution of gas emission can be a significant portion ($\sim$ 30\%) of the total FUV emission, although it does not contribute to the disk heating, since it likely originates in the upper layers of the disk \citep{ingleby09,bergin04,herczeg02,herczeg04,france10}.  In addition, we measure the flux in the strong FUV atomic lines of existing spectra, which our accretion column models do not attempt to reproduce and find that $\sim$ 35\% of the total FUV flux comes from hot lines.  Another contribution necessary for re-producing the short wavelength FUV continuum ($\sim1300-1400$ {\AA}) could  originate in post-shock emission escaping absorption by the pre-shock material.  The post shock spectrum peaks in the soft X-rays but extends into the FUV around 1300 {\AA} \citep{calvet98}.  This emission would be important for disk photoevaporation models as soft X-rays can efficiently heat disk gas \citep{owen10}, but may only play a role when the $\mdot$ drops and the accretion columns become less dense \citep{gorti09b}.  Additional observations are required to determine if this is a viable contribution to the FUV spectrum.

\begin{figure}
\plotone{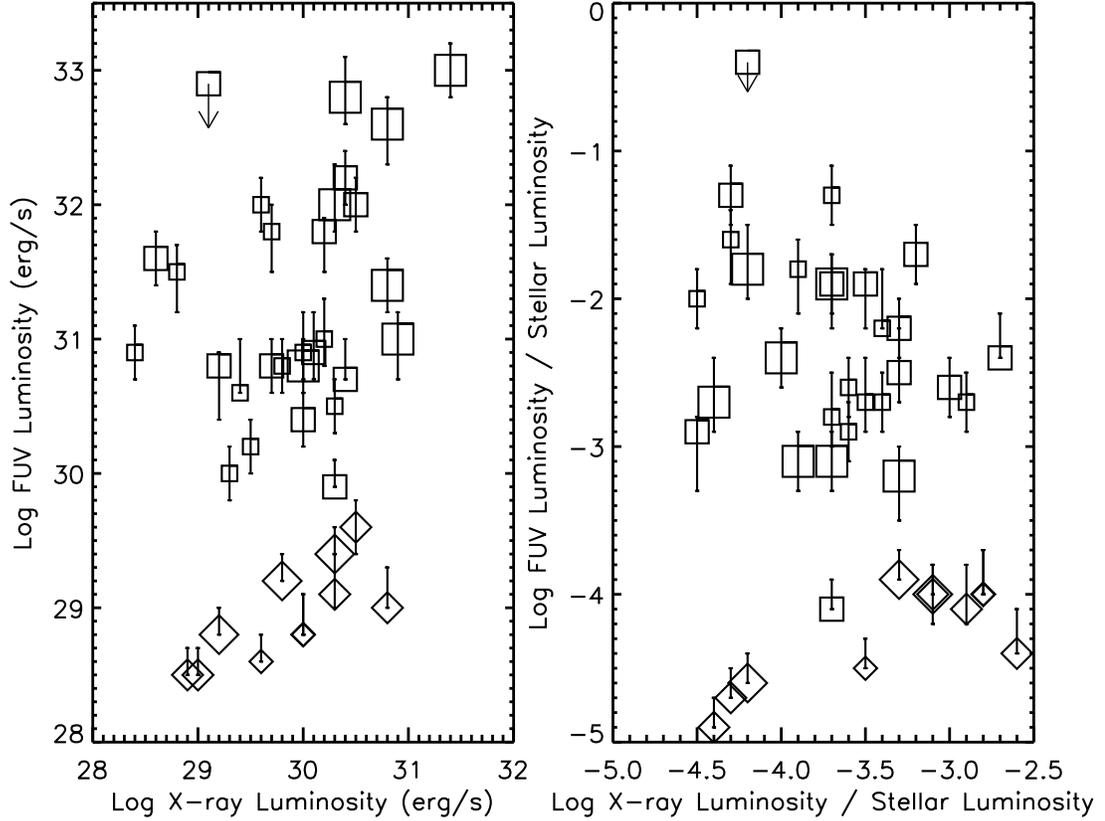}
\caption{\emph{Left:} Comparison of total FUV and X-ray luminosities.  Boxes represent accreting sources and diamonds represent non-accreting sources.  The largest symbols are G stars, medium symbols are K stars and the smallest symbols are M stars.  There is no correlation between FUV and X-ray luminosities among this sample; however, the non-accretors have the lowest FUV luminosities.  The subset of non-accretors has weakly correlated total X-ray and FUV luminosities.  \emph{Right:} Comparison of fractional X-ray and FUV luminosities.  Again, no correlation is observed but non-accretors have the lowest fractional FUV luminosities.  One CTTS, CR Cha, appears in the region populated by non-accretors ($<-3.5$).  We note that this is the only Chamaeleon I source with $A_V=0$.}
\label{fuvvsxray}
\end{figure}

After accretion has ended and the gas has been dissipated at $>$10 Myr, FUV luminosities continue to decline.  This is also observed in magnetically active young stars; all high energy radiation fields decrease as the sources evolve towards and even along the main sequence \citep{ribas05}.  Given this fact, we looked for any relation between the FUV and X-ray emission in the CTTS and non-accretors.  In Figure \ref{fuvvsxray} we distinguish non-accretors (diamonds) from CTTS (squares) and indicate hotter stars by larger symbols.  We see no correlation between FUV and X-ray emission in the CTTS (with Pearson correlation coefficients of 0.3 and -0.2 for total and normalized luminosities, respectively) but we note that the sources with earlier spectral types tend to have the strongest total FUV and X-ray emission, as they are observed to occupy the upper right hand corner in the left panel of Figure \ref{fuvvsxray}.  FUV and X-ray luminosities normalized by the stellar luminosity are indistinguishable by spectral type (right panel of Figure \ref{fuvvsxray}).  We do observe a weak correlation between both total and normalized X-ray and FUV luminosities when considering only the non-accreting sample, with Pearson correlation coefficients of 0.74 and 0.80, respectively.  

\begin{figure}
\plotone{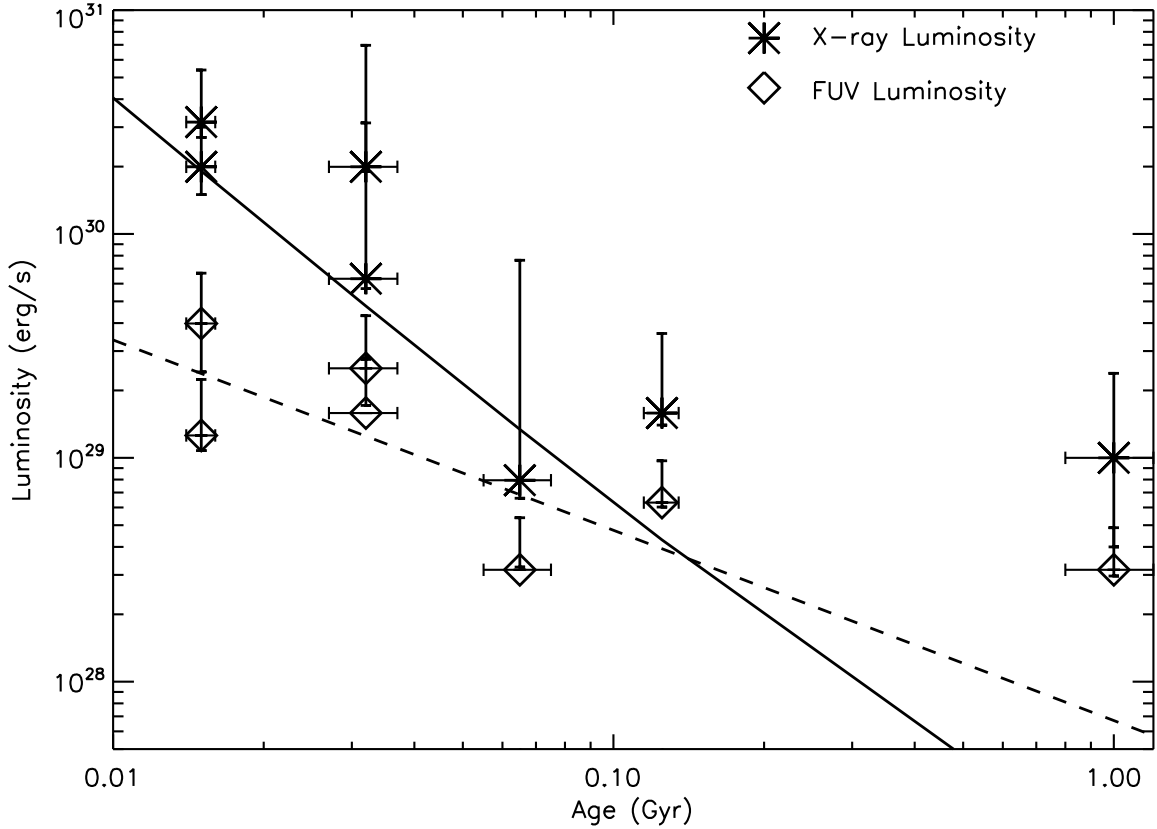}
\caption{The luminosity of young solar analogs compared to power law relations of high energy radiation fields derived in \cite{ribas05}.  The solid line shows the luminosities predicted for the summed 1-- 20 {\AA} and the 20-100 {\AA} flux bins as defined by \cite{ribas05}, approximately the same wavelength region as our 0.2-10 keV observations.  The dashed line shows the slope of the predicted fluxes from the 920--1180 {\AA} wavelength bin but scaled vertically in luminosity to fit the FUV luminosities of our longer wavelength observations.  The slope of the FUV power law relation provides a good fit to our observed FUV luminosities.  The predictions fail to fit the 1 Gyr source in both the X-ray and FUV regimes.}
\label{ribas}
\end{figure}

We compare results from our sample of non-accreting young stars (excluding the low mass TWA sources) to an analysis of solar analogs between 0.1 and 10 Gyr \citep{ribas05}, which showed that high energy fields between 1 and 1200 {\AA} decreased with age, with emission from hotter plasmas diminishing more quickly.  For our sample of non-accreting solar type sources between 15 Myr and 1 Gyr, classified as young suns in the Formation and Evolution of Planetary Systems survey (FEPS; Meyer et al.  2006), we compare the observed luminosities to those predicted assuming the power law relations between flux in a given energy range and age from \cite{ribas05}.  The predicted decay of X-ray luminosities is similar to that which we observe for the youngest of the solar type non-accretors in our sample but fail to re-produce the 1 Gyr source, HD 53143 (see Figure \ref{ribas}).  We also compare the slope of the predicted FUV fluxes between 920--1180 {\AA} to the slope of our observations which cover 1230--1800 {\AA}.  The total predicted luminosities were scaled, due to the use of different wavelength bins, but the slope was held constant.  Again, the younger sources follow the predicted slope but the 1 Gyr source has an unexpectedly high FUV luminosity.  It is unclear why HD 53143 is discrepant in both X-rays and FUV fluxes for its age, with several different age indicators providing consistent values \citep{song00}.

Our young solar analogs extend the radiation evolution observed in \cite{ribas05} to younger ages, and agree with their result that the higher energy radiation fields decline more quickly than the cooler fields.  Combining our results with those of \cite{ribas05}, we see evidence that even past the 10 Myr time period for disk dispersal and planet formation, high energy fields strongly in excess of the solar values persist and may have an effect on evolving disks and planets.

\subsection{The He II $\lambda$1640 and C IV
$\lambda$1549 lines}

Using observations for a limited number of T Tauri stars, \cite{alexander05} have
investigated the origin of the He II $\lambda$1640 and C IV
$\lambda$1549 lines and their relationship to the
continuum flux, responsible for photoevaporation of the gas disk.
In particular, \cite{alexander05} propose that the
ratio of the He II to C IV is a probe of the EUV
radiation field. They base
their conclusion on the fact
that single temperature and density models of 
collisionally ionized plasmas cannot reproduce the observed strengths
of the He II line, from which they infer that the
line  must be produced by radiative recombination.  Since the ionization threshold for He I is 228 {\AA}, they conclude that
the line is a probe of the strength of the radiation field around this 
wavelength, namely, the EUV.

\begin{figure}
\plotone{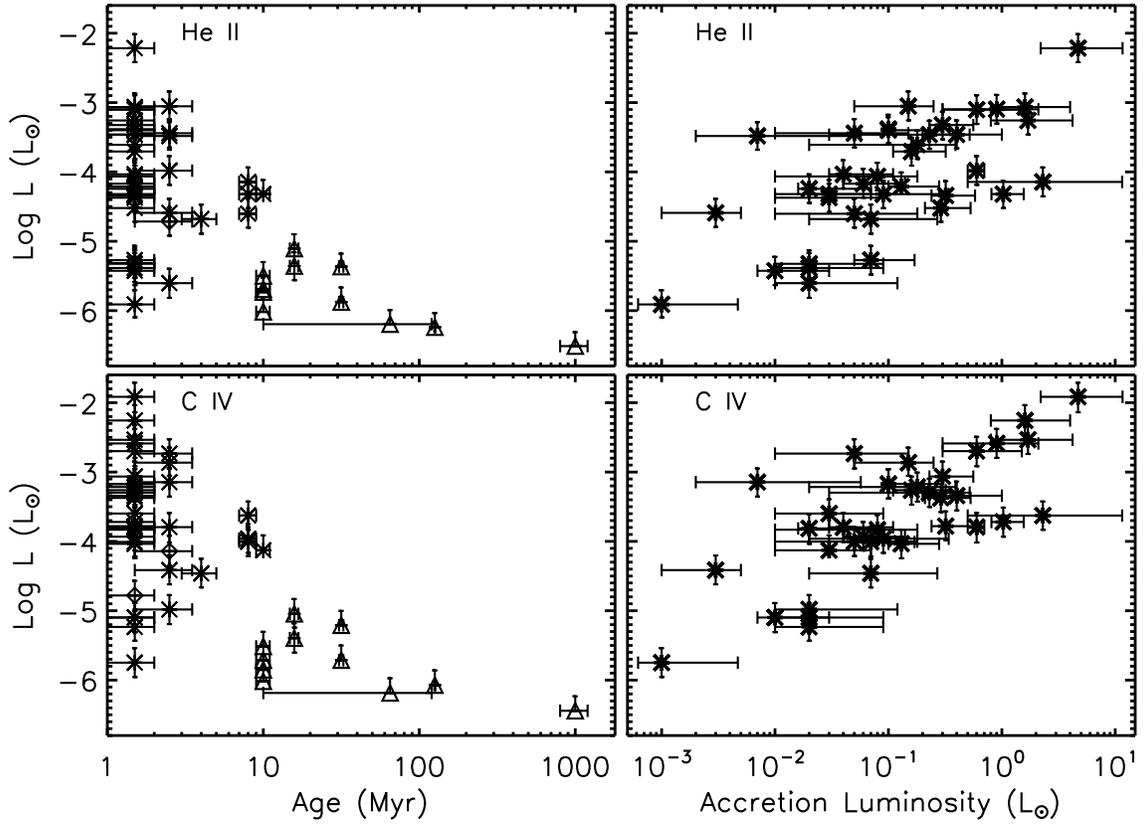}
\caption{\emph{Top Left:} He II $\lambda$1640 {\AA} luminosity vs age.  \emph{Bottom Left:} C IV $\lambda$1549 {\AA} line luminosity vs age.  Symbols are defined as in Figure \ref{fuvvslaccage}.  In both cases the luminosity of the line is observed to decrease with age.  \emph{Top Right:} He II $\lambda$1640 {\AA} luminosity vs $L_{acc}$.  \emph{Bottom Right:} C IV $\lambda$1549 {\AA} line luminosity vs $L_{acc}$.  Both He II and C IV are correlated with $L_{acc}$.}
\label{linevsage}
\end{figure}

Here, we revisit these lines using our larger sample of FUV spectra of T Tauri stars.
We measure the He II and C IV fluxes in our ACS/SBC and STIS samples
of sources with the STIS observations degraded to the low resolution
of ACS/SBC.  As a result of the resolution, we are measuring a blend
of emission lines in each line feature, including both components of
the C IV doublet and numerous \h2 lines present in the vicinity of C
IV and He II.  While adding to the scatter in our measurements, the
dominant line in each region is the He II or the C IV line, with \h2
line contributions $<10\%$.  As discussed in $\S$\ref{fuvlum}, the
largest source of uncertainty comes from the reddening correction of the FUV spectra,
so to estimate errors on the He II and C IV line luminosities we
assume a range of $\pm$0.2 magnitudes in $A_V$.  Figure
\ref{linevsage} shows He II and C IV line luminosities and their
dependence on age and $L_{acc}$.  Both He II and C IV line
luminosities are observed to decrease with age with the lowest values
obtained for non-accretors.  We find strong correlations between line
luminosities and $L_{acc}$, with Pearson coefficients $\sim$ 0.84 and
0.88 for He II and C IV, respectively.  The
strength of both lines is clearly related to accretion, as was previously
found by 
\cite{calvet04} and \cite{johnskrull00}.

\begin{figure}
\plotone{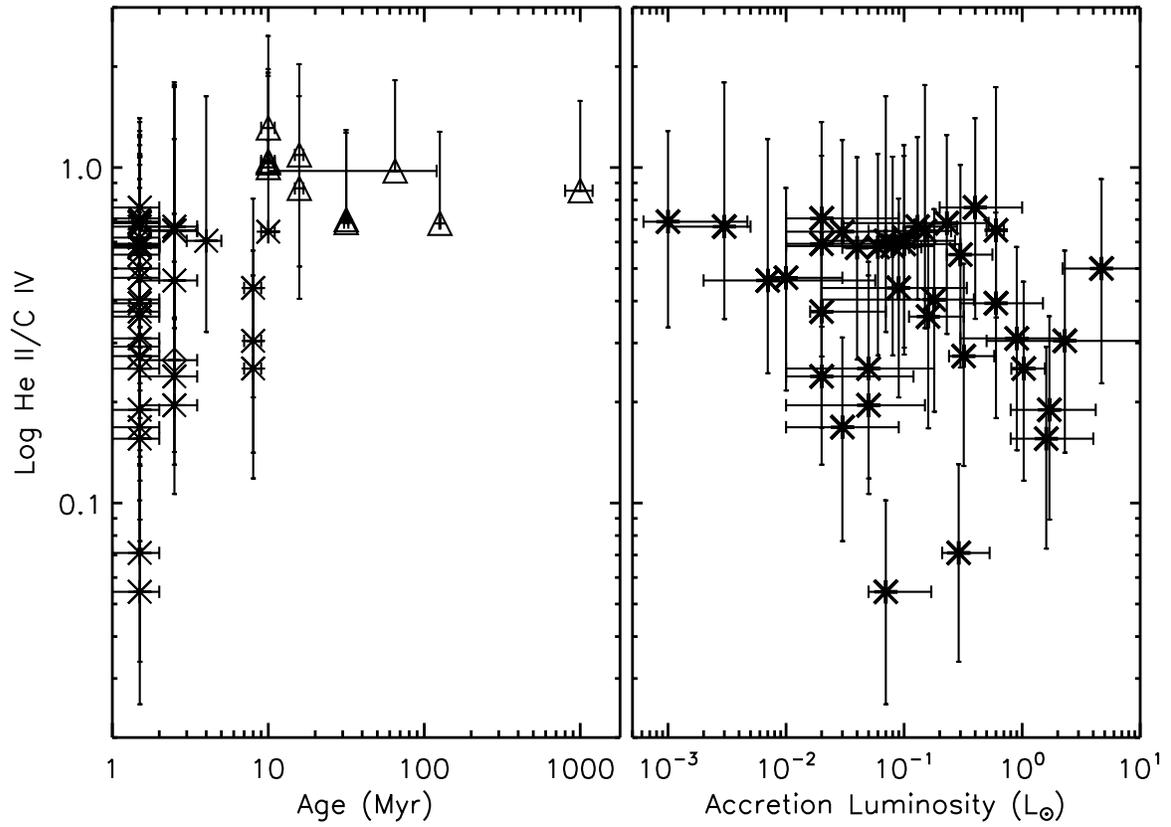}
\caption{Ratio of He II to C IV as a function of age (left) and accretion luminosity (right).  Symbols are defined as in Figure \ref{fuvvslaccage}.  We see that the non-accretors clearly have a higher value of He II/ C IV, approximately 1, while the CTTS have He II/ C IV $<$ 1; however, we see no clear trend with accretion luminosity.}
\label{ratiovsagelacc}
\end{figure}

In 
the left panel of Figure
\ref{ratiovsagelacc} we show the He II to C IV ratio for the CTTS and non-accretors
of our sample as a function of age. The value of the ratio is higher in non-accretors than in the CTTS,
confirming the trend found by \cite{alexander05}.
These authors suggest that this trend indicates that the
ionizing flux is not powered by accretion, since the
ratio would be expected to decrease if that were the case.
We suggest, instead, that the difference between the He II to C IV
ratio between the CTTS and the non-accretors arises from the different
formation mechanisms of the lines in each type of star.
The He II $\lambda$1640 and C IV
$\lambda$1549 lines in non-accretors must form in the transition
region, as they do in the sun and in magnetically active stars, indicated
by the similarity of the ratio in the two classes of stars \citep{hartmann1979,byrne1989}.
In contrast, the correlations between line luminosity and
$L_{acc}$ found in CTTS indicate
that these lines must form in the accretion flows
or at least in regions powered by accretion
energy. However, despite the line luminosity dependence
on $L_{acc}$, we find no correlation
between the He II to C IV ratio and $L_{acc}$, as shown in the right panel
of Figure \ref{ratiovsagelacc}. Studies of the optical
and infrared lines of He I and He II in CTTS
indicate that the lines have multiple components
arising in the accretion shocks, the magnetospheric infall
regions, and the winds \citep{beristain2001,edwards2006}.
No comparable analysis of the C IV line has been done, but it is
also likely that this line has contributions from the accretion
shock and from the infall region, given its large velocity
width \citep{ardila2002}.
This diversity may be responsible for the
large dispersion of the He II to C IV ratios in CTTS.
We therefore caution against using
the He II to C IV ratio as an indicator
of EUV emission without a careful treatment of accretion related emission in these lines.

Regardless of the effectiveness of the He II to C IV line ratio in
tracing the EUV emission, strong EUV emission is likely present in the
high energy spectrum of T Tauri stars.  \cite{ribas05} showed that EUV
emission decreases with age, or increases with the spin of the source.
They also showed that younger sources have enhanced EUV line emission,
in particular strong He II and Fe lines around 300 - 350 {\AA}.
Whether the EUV emission continues to increase at $<$10 Myr like the
FUV emission or flattens and remains constant like the X-ray emission
is not clear, but in either case the EUV emission is expected to be
strong.  In our model of the accretion column which contributes emission to the FUV spectrum (\S\ref{xray_uv}), we find that  the EUV luminosity produced in the shock is $\sim10^{31}\;\rm{erg s^{-1}}$.  In the current model, only $\sim$2\% of this emission escapes the high density material which is still accreting; however, recent work has shown that varying accretion properties, like the accretion rate or infall velocity, affects the amount of shock emission which can be observed \citep{sacco08}.  Future models of the emission from the accretion column which take this into account will provide better estimates of the EUV emission which is available for photoevaporation.

\subsection{High Energy Radiation Fields and Photoevaporation}
\label{he}

In contrast to early photoevaporation models in which the primary mechanism for dispersing the disk utilized EUV radiation from the star producing winds with mass loss rates of $\sim10^{-10}\;\msunyr$ \citep{hollenbach94}, recent models stress that either FUV or X-ray radiation fields are significantly more important in dispersing the disk with mass loss rates 100 times higher than in the pure EUV case.  While in agreement on the values of the mass loss rate, \cite{owen10} achieve these high rates with X-ray radiation while \cite{gorti09b} find the FUV radiation drives the photoevaporation.  With the large range of ages in our X-ray and FUV samples, we explore the photoevaporation theories in the context of our observations.

Our results agree with previous findings by \cite{preibisch05} and \cite{mercer09} showing that the X-ray luminosity remains roughly constant during the first 10 Myr.  Regarding models of X-ray photoevaporation,  \cite{owen10} use $L_X=2\times10^{30}\;\rm{erg s^{-1}}$ for $h\nu>0.1$ keV as their X-ray luminosity.  We find similar, though consistently lower, X-ray luminosities for the typical CTTS mass sources in our sample.  \cite{owen10} note that their input X-ray luminosity is large and that further analyses will include lower X-ray luminosities.

An important new result in this work is the evolution of the FUV radiation, where we see that the luminosity between 1230 and 1800 {\AA} clearly drops between 1 and 10 Myr.  \cite{gorti09b} incorporate the evolution of FUV luminosity in their models for disk dispersal, assuming that $\mdot$ evolves viscously and the FUV luminosity is 0.4 $L_{acc}$.  In Figure \ref{fuvvslaccage} we showed the evolution of the FUV luminosity along with the predicted FUV emission produced in an accretion column from a viscously evolving disk onto a 0.8$\msun$ star.  We found that the predicted FUV accretion emission is lower than the observed FUV fluxes, although, within an order of magnitude (assuming typical mass CTTS have $L_{acc}<0.5\;\lsun$) and we discussed origins for the remaining FUV emission in $\S$\ref{xray_uv}.  The FUV fluxes used by \cite{gorti09b} are similar to those we observe, but we found that $\sim$1/3 of this emission may be produced by \h2 in the circumstellar disk (Ingleby et al. 2011) and therefore not directly irradiating the gas in the disk.

Our observations support the requirements of both the X-ray and the FUV dominated photoevaporation models of \cite{owen10} and \cite{gorti09}; however, several issues remain unresolved.  For example, circumstellar disks of gas and dust have been observed around much older sources, including the 10 Myr CTTS TW Hya which maintains a 0.06 $\msun$ disk, albeit a transitional disk \citep{calvet02,najita10}.  Full disks of gas and dust are also observed around the 7--10 Myr CTTS in 25 Ori \citep{ingleby09,hernandez07}.  According to \cite{owen10}, the continued existence of full disks at late stages could be explained by the range of X-ray luminosities observed, with lower X-ray luminosities increasing the disk lifetime and, alternatively, the presence of young WTTS may be due to high X-ray luminosities in those sources.  In our 25 Ori observations in which $\sim$90\% are WTTS, we find that the median X-ray luminosity (log $L_X$) is 29.7 erg $\rm{s^{-1}}$ and the four accreting sources have only marginally lower or, in one case, even higher X-ray luminosities.  With similar X-ray luminosities to the WTTS, these accreting sources should have cleared their inner gas disks in similar timescales.  If we look at the X-ray luminosities of the WTTS in the 2.5 Myr Chamaeleon region, we find a median log $L_X=29.0$ $\rm{s^{-1}}$, lower than all of the 25 Ori accretors, indicating that if these X-ray luminosities may disperse an inner gas disk in $\sim$2.5 Myr, they would clearly be gone by 10 Myr.  An additional parameter, besides a range of X-ray luminosities, is necessary to explain the diversity of disks observed at a given age.

When including the time evolution of the FUV radiation fields and their dependence on disk properties, like $\mdot$, \cite{gorti09b} find FUV fields evolving on a range of timescales related to differences in the accretion properties.  Therefore, photoevaporation dominated by evolving FUV fields may disperse disks in a range of timescales.  With FUV driven photoevaporation and evolving FUV fields, the diversity of disks observed at any age may be expected.  This appears to be a plausible scenario given the large range of observed FUV luminosities and inferred $\mdot$'s.  Again, the full disks in the 25 Ori association present an obstacle to the theory.  The observed FUV luminosities of the 25 Ori accretors are $\sim$ 2 orders of magnitude higher than the WTTS in the 10 Myr TWA association, and given our results and those of \cite{ribas05}, were likely stronger at 1 Myr.  These FUV luminosities should be high enough to photoevaporate the gas disk, if FUV emission was responsible for the gas depletion in the TWA WTTS.  Possible explanations for the remaining full disks in 25 Ori, in the context of both X-ray and FUV photoevaporation theories, include low disk viscosities or high initial disk masses which increase the time for disk dispersal \citep{owen10}.

\subsection{FUV Radiation and Dust Evolution}  

So far, we have focused on the gas in circumstellar disks; however, it is unclear if gas and dust evolution proceed concurrently or independently \citep{pascucci06}.  \cite{alexander07} modeled the affect of photoevaporation on the gas and dust in the circumstellar disk and found that the removal of gas from the disk has significant effects on the dust.  When photoevaporation creates a gap in the gas disk, the inner disk dust is also removed.  Also, pressure gradients formed when gas is removed cause the dust to migrate and may enhance grain growth.  For these reasons, we may expect to see dust evolution dependent on the strength of the photoevaporative fields.  With constant X-ray fields throughout disk evolution, we focus on the accretion related FUV radiation fields.

\begin{figure}
\plotone{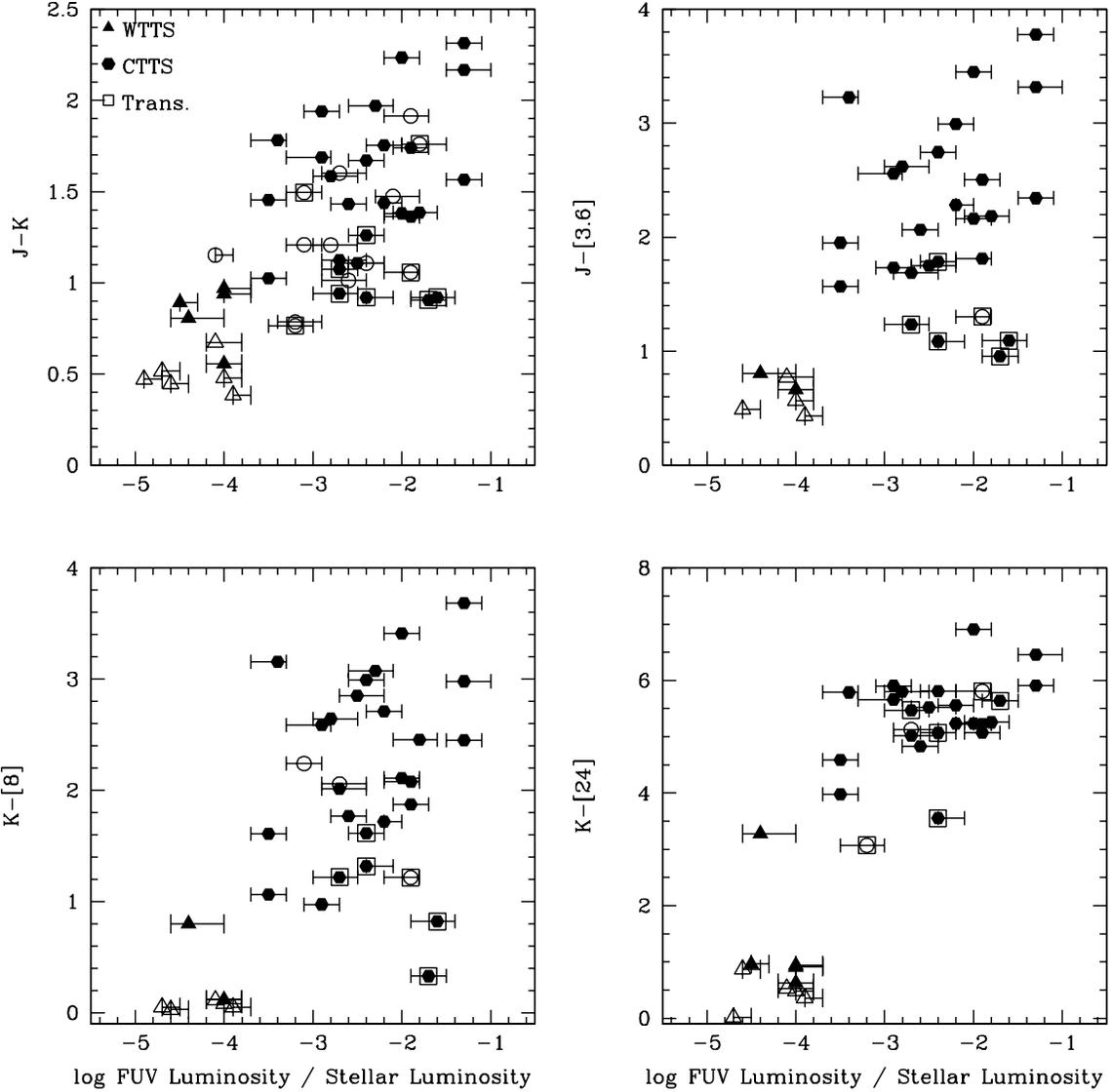}
\caption{FUV radiation and tracers of disk evolution.  We compare the FUV luminosities to several indicators of evolution in the disk.  [3.6], [8], and [24] represent IRAC and MIPS photometry at 3.6, 8, and 24 $\mu$m respectively.  Open symbols are sources with spectral types earlier than K5 and closed symbols are spectral types K5 and later.  $J$-[3.6], $K$-[8], and $K$-[24] probe regions progressively further in the disk.  We see only tentative correlations between high energy emission and dust evolution tracers, but this is likely due to both quantities being dependent on accretion rate.  We do see that the full and transitional or pre-transitional disks (boxed symbols) have similar FUV luminosities, indicating that the FUV radiation is not very significant in dust evolution.}
\label{disk}
\end{figure}

 \cite{fedele10} find that dust dissipation and accretion are characterized by different timescales and to test this further, we compare FUV luminosities to 2MASS $J,H,K$ \citep{skrutskie06}, IRAC (\emph{Spitzer} InfraRed Array Camera), and MIPS (\emph{Spitzer} Multiband Imaging Photometer) colors.  IRAC and MIPS photometry, along with disk classifications for our source sample, were taken from the literature: Taurus \citep{luhman10}, Chamaeleon I \citep{luhman08}, Chamaeleon II \citep{alcala08}, 25 Ori \citep{hernandez07}, TWA \citep{low05,hartmann05}, and debris disks \citep{carpenter08,chen05}.  Figure \ref{disk} shows the relations between FUV luminosity and inner disk dust tracers.  At $\lambda<24\;\mu$m we are tracing the dust in the inner regions of the disk; as shown in \cite{dalessio06}, 60\% of the 24 $\mu$m emission is coming from inside 10 AU when the dust has not settled towards the midplane.  The percentage of emission coming from within 10 AU increases when observing at the shorter IRAC wavelengths, or when observing a disk in which some degree of dust settling has occurred.    

Among the CTTS (circles) in Figure \ref{disk}, we find tentative correlations between FUV luminosity and each of the disk evolution tracers, with the most convincing correlation in $J-K$.  However, we are likely observing two quantities which correlate with $L_{acc}$.  We showed in Figure \ref{fuvvslaccage} that the sources with highest FUV luminosities are also the strongest accretors.  Disk properties also depend on the accretion rate through the disk and the IR spectral energy distributions reflect this; models predict that lower $\mdot$'s will produce bluer IR colors and that the IR spectrum becomes increasingly flatter as $\mdot$ increases \citep{dalessio06,espaillat09}.  Therefore, it is likely that these apparent trends are a result of accretion affecting both the FUV luminosities and the tracers of dust in the disk  instead of a consequence of direct processing of the dust by FUV radiation.  This seems to indicate that if it is the FUV that is responsible for gas dispersal, the gas and dust are evolving independently.  Supporting this conclusion, the transitional \citep{calvet02,calvet05,kim09} and pre-transitional \citep{espaillat08,espaillat10} disks, the sources likely to have undergone dust processing and evolution, have the same FUV luminosities as the full disks, indicating that the FUV is not likely having a strong effect on the dust.

\section{Summary and Conclusions}
\label{summary}
We present new observations of T Tauri stars in the Chamaeleon I and II and the 25 Ori star forming regions.  Included are \emph{Chandra} X-ray and ACS/SBC FUV observations which provide constraints on the photoevaporative fluxes present in young stars.  Combined with previously published X-ray and FUV observations, we explored the evolution of high energy radiation fields with the following results:
\begin{itemize}
\item 
With the addition of $\sim$40 sources with X-ray observations in the 7--10 Myr 25 Ori star forming association, a significant addition to the $\sim$25 previously observed 6--10 Myr T Tauri stars (above the substellar limit), we confirm results that there is no decay in the X-ray luminosity between 1 and 10 Myr.  With consistently strong X-ray radiation fields, disk photoevaporation, if driven by X-rays, should proceed on short timescales.  However, observations of sources with remaining disk gas emission and strong X-ray luminosities at 7--10 Myr and previous observations of TW Hya (a source with a transitional disk and strong X-ray luminosity) still cannot be explained by X-ray photoevaporation.

\item
We show, for the first time, the observed evolution of the FUV radiation fields in pre-main sequence stars from 1 Myr to 1 Gyr.  We find that the FUV luminosity decline can be explained, in part, by the decline of the accretion rate onto the star but note that some FUV emission must come from an additional source which is likely gas in the inner disk regions.  If FUV drives photoevaporation of the gas disk, the diversity of disks observed at 1 Myr can be understood by the large range of $\mdot$'s observed at 1 Myr.  However, gas disks at 7--10 Myr require additional explanation, as viscous $\mdot$ evolution predicts that these sources should have had strong FUV radiation in the past, so disk dispersal by FUV radiation should have proceeded quickly.

\item
We show that the power law relations for the decay of high energy radiation fields in solar type active stars described by \cite{ribas05} extend to the significantly younger sample discussed here.  With both the X-ray luminosities and the slope of the FUV relation accurately predicted, we expect that the difficult to observe EUV radiation fields will similarly obey the power law trends.  This result tells us that strong radiation fields, necessary to remove the gas, are present throughout the disk dispersal phase.

\item
We find no correlation between FUV luminosities and tracers of dust evolution in the disk and furthermore, we find that sources with significant dust processing, the transitional disks, have similar FUV luminosities to full disk sources.  Therefore, if FUV radiation is the key emission field in gas evolution, it is not driving the dust depletion.  This would indicate that gas and dust evolution occur independently, though interestingly on similar timescales.
\end{itemize}

\section{Acknowledgments}
We thank Kevin Luhman for assistance in identifying possible new T Tauri stars in Chamaeleon I.  We also thank Jennifer Blum and Jun Ji for their help with X-ray data reduction and analysis techniques.  We thank Gregory Herczeg for his assistance with reducing the ACS Chamaeleon data.  Support for this work was provided by NASA through Chandra Awards GO8-9028X and GO8-9029X issued by the Chandra  X-ray Observatory Center, which is operated by the Smithsonian Astrophysical Observatory for and on behalf of NASA under contract NAS8-03060. Support for this work was also provided by NASA through grant Nos. GO-08317, GO-09081, GO-09374, and GO-11145 from the Space Telescope Science Institute, which is operated by AURA, Inc., under NASA contract NAS 5-26555.  C.E. was supported by the National Science Foundation under Award No. 0901947.  This publication makes use of data products from the Two Micron All Sky Survey, which is a joint project of the University of Massachusetts and the Infrared Processing and Analysis Center/California Institute of Technology, funded by the National Aeronautics and Space Administration and the National Science Foundation.




\begin{deluxetable}{llccccccc}

\tablewidth{0pt}
\tablecaption{Properties of Sample
\label{resultstable}}
\tablehead{
\colhead{Source} & \colhead{Alt. Name} &\colhead{SpT}& \colhead{$\rm{L_X}$} & \colhead{$\rm{L_{FUV}}$}  & \colhead{$\rm{ L_{\ast}}$ }& \colhead{$\rm{L_{acc}}$}& \colhead{Reference$^{a,b}$}\\
\colhead{ }&\colhead{ }   &\colhead{ }    &\colhead{($\lsun$) } &\colhead{ ($\lsun$)}      & \colhead{($\lsun$)} & \colhead{ ($\lsun$)}  \\}
\startdata
Taurus\\
AA Tau &&M0&$\it{1.3\times10^{-4}}$     &0.016&\emph{1.0}&\emph{0.13}&1, 15\\
\bf{BP Tau}&&K7&$\it{4.1\times10^{-4}}$&0.016&\emph{1.3}&\emph{0.23}&1, 15\\
CI Tau  &&K6&$\it{4.1\times10^{-5}}$      &$1.6\times10^{-3}$&\emph{1.3}&\emph{0.47}&1, 15\\
DE Tau &&M1&--&0.013&\emph{1.3}      &\emph{0.16}&1\\
DL Tau &&K7&--&$4.1\times10^{-3}$&\emph{1.0}       &\emph{0.32}&1\\
\bf{DM Tau}  &&M1&$\it{1.6\times10^{-5}}$&$8.3\times10^{-3}$&\emph{0.33}&\emph{0.08}&1, 15\\
DN Tau &&M0&$\it{4.1\times10^{-4}}$    &$2.6\times10^{-3}$&\emph{1.3}&\emph{0.04}&1, 15\\ 
DO Tau &&M0&--&0.066&\emph{1.3      }&\emph{0.29}&1\\
DP Tau &&M0&$\it{6.6\times10^{-6}}$    &$2.1\times10^{-3}$&\emph{0.21}&\emph{0.01}&1, 15\\
DR Tau &&K7&--&$8.3\times10^{-3}$&\emph{1.6}&\emph{1.03}&1\\
FM Tau &&M0&$\it{1.0\times10^{-4}}$&0.026&\emph{0.52}&\emph{0.30}&1, 15\\
FP Tau &&M3&--&$1.3\times10^{-4}$&\emph{0.41}&\emph{0.001}&1\\
GK Tau &&M0&$\it{2.6\times10^{-4}}$&$2.1\times10^{-3}$&\emph{1.3}&\emph{0.06}&1, 15\\
\bf{GM Aur}   &&K3&--&0.016&\emph{1.3}&\emph{0.18}&1\\
HN Tau A$^{\ast}$ &&K5&$\it{1.0\times10^{-5}}$&0.010&\emph{0.21}&\emph{0.02}&1, 15\\
HN Tau B$^{\ast}$&&M4&--&$2.6\times10^{-4}$&\emph{0.033}&--&1 \\
IP Tau   &&M0&$\it{1.6\times10^{-4}}$& $1.6\times10^{-3}$&\emph{0.66}&\emph{0.02}&1, 17\\
\bf{LkCa 15} &&K5&--&$4.1\times10^{-3}$&\emph{1.0}&\emph{0.03}&1\\
\bf{RY Tau}   &&G1&$\it{6.6\times10^{-4}}$&0.16&\emph{10.4}&\emph{1.6}&1, 15\\
\bf{SU Aur}   &&G1&$\it{1.6\times10^{-3}}$&$6.6\times10^{-3}$&\emph{8.3}&\emph{0.10}&1, 15 \\
\bf{T Tau}      &&G6&$\it{1.6\times10^{-3}}$&0.10&\emph{8.3}&\emph{0.90}&1, 15\\
UZ Tau A$^{\ast}$ &&M1&--&$1.3\times10^{-4}$&\emph{0.33}&\emph{0.02}&1\\
UZ Tau B$^{\ast}$ &&M2&$\it{8.3\times10^{-5}}$&$4.1\times10^{-4}$&\emph{0.33}&\emph{0.02}&1, 15\\
\hline
Orion\\
\bf{CO Ori}    &&G0&--&0.16&\emph{20.8}&\emph{1.7}&1\\
\bf{EZ Ori}     &&G3&--&0.10&\emph{6.7}&\emph{0.10}&1 \\
\bf{GW Ori}   &&G0&$\it{6.6\times10^{-3}}$&0.26&\emph{65.7}&\emph{4.7}&1, 16\\
\bf{P 2441}    &&F9&--&$6.6\times10^{-3}$&\emph{10.4}&\emph{0.4}&1\\
\bf{V1044 Ori}&&G2&--&0.016&\emph{6.7}&\emph{0.6}&1 \\
\hline
Chamaeleon I\\

CHXR 8    &     &K2&$2.1\times10^{-5}$               &--             &--                   &--&13\\
CHXR 30A&    &K8&$1.0\times10^{-4}$             &--             &\emph{1.3}&0&2, 3\\
CHXR 30B&    &M1&$8.3\times10^{-6}$              &--             &\emph{0.21}&0&2, 3\\
CR Cha     &T8&K2&$5.2\times10^{-4}$               &$2.1\times10^{-4}$       &\emph{2.6}&0.02&3\\
CS Cha      &T11&K6&$\it{8.3\times10^{-4}}$&0.026       &\emph{1.3}&0.01&2, 3, 18\\
DI Cha        &T26&G2&$\it{5.2\times10^{-4}}$&0.026       &\emph{13.1}&0.15$^{\ast}$&3, 17\\
FK Cha       &T29&K6&--                   &$<$5.2&\emph{2.6}&1.2&3\\
FL Cha      &T35&K8&$3.3\times10^{-5}$              &$<$0.21&\emph{0.52}&\emph{0.04}&2, 3, 8\\
FO Cha       &T47&M2&$2.1\times10^{-5}$              &--            &\emph{0.41}&\emph{0.01}&2, 3, 8\\
HN 7            &&M5&$8.3\times10^{-6}$              &--             &\emph{0.066}  &0&2, 3\\
HN 12W      &&M6&$8.3\times10^{-6}$               &--            &\emph{0.066}&0&2, 3\\
ISO 86         &&--&$2.1\times10^{-5}$              &--&--&--\\
ISO 91        &&M3&$6.6\times10^{-4}$              &--            &\emph{0.26}&--&2, 3\\
ISO 97        &&M1& $3.3\times10^{-5}$              &--            &--&--&2\\
ISO 196      &&K7&$1.0\times10^{-5}$               &--            &--&--\\
SY Cha      &T4&K6&$1.0\times10^{-4}$               &--            &\emph{0.52}&0.002&2, 3\\
SZ Cha       &T6&K0&$2.6\times10^{-4}$               &$1.6\times10^{-3}$      &\emph{2.1}&--&3\\
T25              &SZ 18&M3&--                    &$<$$1.3\times10^{-3}$&\emph{0.26}&0.01&2, 3\\
T54              &&G8&$\it{2.1\times10^{-3}}$&$2.6\times10^{-3}$       &\emph{4.1}&0.58$^{\ast}$&3, 18\\
T56              &SZ 45&M1&$\it{5.2\times10^{-4}}$&$8.3\times10^{-4}$        &\emph{0.41}&0.003&2, 3, 17\\
TW Cha       &T7&K8&$1.0\times10^{-4}$            &--              &\emph{0.52}&0.01&2, 8\\
VW Cha       &T31&K8&$\it{6.6\times10^{-4}}$&0.041       &\emph{3.3}&0.05&2, 3, 18\\
\hline
Chamaeleon II\\
BF Cha&SZ 54&K5&$\it{2.6\times10^{-4}}$ &$6.6\times10^{-4}$&\emph{2.1}&0.07&4, 17\\
BM Cha&SZ 61&K5&--&$<1.6\times10^{-4}$&\emph{1.3}&\emph{0.64}&4\\

\hline
25 Ori\\
CVSO 24&OB1a 499&M2& $8.3\times10^{-5}$&--&\emph{0.16}&\emph{0.01}&5, 10, 12\\
CVSO 29&OB1a 931&M3& $1.0\times10^{-4}$&--&\emph{0.26}&--&5, 10, 12\\
CVSO 34&CVSO 221&M0&$4.1\times10^{-4}$&--&\emph{0.21}&0&5, 9\\
CVSO 35&OB1a 1192&K7&$3.3\times10^{-4}$&$2.1\times10^{-3}$&\emph{0.66}&0.05&5, 10, 12\\
CVSO 36&OB1a 1222&M2&$3.3\times10^{-4}$&--&\emph{0.21}&\emph{0.03}&5, 10, 12\\
CVSO 38&OB1a 1317&M2&$4.1\times10^{-4}$&--&\emph{0.16}&\emph{0.02}&5, 10, 12\\
CVSO 39&OB1a 1418&M2&$2.1\times10^{-4}$&--&\emph{0.33}&\emph{0.01}&5, 12\\
CVSO 43&OB1a 1716&M2&$1.3\times10^{-4}$&--&\emph{0.26}&\emph{0.02}&5, 12\\
CVSO 200&OB1a 489&M3&$4.1\times10^{-4}$&--&0.21&--&11, 12\\
CVSO 206&OB1a 776&K7&$1.3\times10^{-4}$&$1.6\times10^{-3}$&0.26&2.3&10, 11, 12\\
CVSO 207&OB1a 864&K6&$4.1\times10^{-4}$&--&0.66&--&10, 11, 12\\
CVSO 211&OB1a 971&K4&$3.3\times10^{-4}$&--&0.52&--&10, 11, 12\\
CVSO 214&OB1a 1022&K2&$2.1\times10^{-4}$&--&0.66&--&10, 11, 12\\
CVSO 217&OB1a 1121&M1&$2.1\times10^{-4}$&--&$>$0.26&--&11, 12\\
CVSO 218&OB1a 1126&M4&$1.3\times10^{-4}$&--&0.13&--&10, 11, 12\\
CVSO 223&OB1a 1160&M3&$2.1\times10^{-4}$&--&0.21&--&10, 11, 12\\
CVSO 224&OB1a 1200&M4&$5.2\times10^{-5}$&$2.6\times10^{-4}$&0.13&\emph{7$\times\it{10}^{-4}$}&6, 10, 11, 12\\
CVSO 225&OB1a 1235&M4&$2.1\times10^{-4}$&--&0.26&--&10, 11, 12\\
CVSO 228&OB1a 1316&M2&$8.3\times10^{-5}$&--&0.16&--&10, 11, 12\\
CVSO 230&OB1a 1326&M1&$2.6\times10^{-4}$&--&0.33&--&10, 11, 12\\
CVSO 233&OB1a 1417&M3&$1.3\times10^{-4}$&--&0.052&--&10, 11, 12\\
CVSO 235&OB1a 1548&M3&$2.1\times10^{-4}$&--&0.16&--&10, 11, 12\\
OB1a 532&&M3&$2.6\times10^{-4}$&--&0.26&--&11, 12\\
OB1a 735&&M5&$3.3\times10^{-5}$&--&$>$0.21&--&11, 12\\
OB1a 834&&M2&$8.3\times10^{-5}$&--&0.16&--&11, 12\\
OB1a 841&&M3&$2.6\times10^{-4}$&--&0.16&--&11, 12\\
OB1a 849&&M5&$8.3\times10^{-5}$&--&0.16&--&11, 12\\
OB1a 948&&M3&$1.0\times10^{-4}$&--&0.13&--&11, 12\\
OB1a 981&&K3&$5.2\times10^{-4}$&--&0.83&--&11, 12\\
OB1a 1088&&K6&$5.2\times10^{-4}$&--&$>$0.52&--&11, 12\\
OB1a 1253&&M4&$8.3\times10^{-5}$&--&0.10&--&11, 12\\
OB1a 1586&&M5&$5.2\times10^{-5}$&--&0.066&--&11, 12\\
OB1a 1626&&K8&$4.1\times10^{-4}$&--&$>$0.83&--&11, 12\\
OB1a 1630&&M2&$6.6\times10^{-5}$&$1.0\times10^{-3}$&0.16&0.09&11, 12\\
OB1a 1663&&M5&$4.1\times10^{-5}$&--&0.10&--&11, 12\\
OB1a 1755&&M3&$1.0\times10^{-4}$&--&0.21&--&11, 12\\
RX J0526.7+0143&&K0&$1.0\times10^{-3}$&--&--&--&14\\
J05244498+0159465&&K6&$8.3\times10^{-4}$&--&0.66&--&11\\
J05260639+0137116&&M2&$6.6\times10^{-5}$&--&0.041&--&11\\
J05263125+0141492&&M4&$5.2\times10^{-5}$&--&0.033&--&11\\
\hline
TWA\\
\bf{TW Hya}  &&K7&$\it{6.6\times10^{-4}}$&$1.3\times10^{-3}$&\emph{0.32}&0.03&1, 16\\
HD 98800    &&K5&$\it{1.6\times10^{-3}}$&$2.6\times10^{-5}$&\emph{0.66}&0&1, 16\\
TWA 7           &&M1&$\it{1.0\times10^{-4}}$&$1.0\times10^{-5}$&\emph{0.32}&0&1, 16\\
TWA 13A      &&M1&$\it{2.6\times10^{-4}}$&$1.6\times10^{-5}$&\emph{0.16}&0&1, 16\\
TWA 13B      &&M1&$\it{2.6\times10^{-4}}$&$1.6\times10^{-5}$&\emph{0.16}&0&1, 16\\
\hline
Field\\
HD 12039    &&G4&$\it{1.6\times10^{-4}}$&$4.1\times10^{-5}$&\emph{0.33}&0&1, 16\\
HD 202917  &&G5&$\it{5.2\times10^{-4}}$&$6.6\times10^{-5}$&\emph{0.66}&0&1, 16\\
HD 53143    &&K1&$\it{2.6\times10^{-5}}$&$8.3\times10^{-6}$&\emph{0.66}&0&7, 16\\
HD 61005    &&G8&$\it{4.1\times10^{-5}}$&$1.6\times10^{-5}$&\emph{0.66}&0&1, 19\\
HD 92945    &&K2&$\it{2.1\times10^{-5}}$& $8.3\times10^{-6}$&\emph{0.41}&0&1, 16\\
MML 28        &&K2&$\it{5.2\times10^{-4}}$&$3.3\times10^{-5}$&\emph{0.41}&0&1, 16\\
MML 36        &&K5&$\it{8.3\times10^{-4}}$&$1.0\times10^{-4}$&\emph{1.0}&0&1, 16\\
\enddata
\tablecomments{\\
Values in italics were taken from the literature while those in regular font were derived in this work.  Sources in bold were observed with STIS rather than ACS/SBC.\\
$^a$Spectral type, stellar, and accretion luminosity references.  1) \cite{ingleby09} and references therein, 2) \cite{luhman04}, 3) \cite{luhman07}, 4) \cite{spezzi08}, 5) \cite{calvet05}, 6) \cite{espaillat08}, 7) \cite{kalas06}, 8) \cite{nguyen09}, 9) \cite{briceno05}, 10) \cite{briceno07}, 11) Brice{\~n}o et a. (2010), 12) \cite{hernandez07}, 13) \cite{huenemoerder94}, 14) \cite{alcala96}\\
$^b$X-ray luminosity references.  15) \cite{gudel07}, 16) \cite{voges99}, 17) \cite{voges00}, 18) \cite{feigelson93}, 19) \cite{wichmann03}\\
$^{\ast}$ $L_{acc}$'s for T26 and T54 were calculated from the $U$ band excess, whereas the remaining Chamaeleon sources have $L_{acc}$ calculated using H$\alpha$ 10\% equivalent widths observed here or taken from the literature.\\}
\end{deluxetable}

\begin{deluxetable}{llcccccc}
\tablewidth{0pt}
\tablecaption{X-ray Detections: Possible and Confirmed T Tauri Members
\label{tabxraynew}}
\tablehead{
\colhead{2MASS Source} & \colhead{SpT}   & \colhead{EW H$\alpha$} & \colhead{EW Li 6706 {\AA}} &\colhead{log $\rm{ L_X}$}  \\}
\startdata
05243490+0154207  &K4--K5&-0.46&0.10&29.0\\
05244265+0154116 &G9--K0&1.54&0.32&--\\
05244320+0200355&--&--&--&-- \\
05251394+0143313&--&--&--&--\\
05265823+0136078&--&--&--&-- \\
05270173+0139157  &K6&-0.39&0.17&29.3\\
10575375-7724495&--&--&--&--\\
\enddata
\tablecomments{Spectral types were determined using the Ca I absorption lines.  With the early spectral type, additional membership confirmation is needed for 2MASS 05244265+0154116 as Li is not a good indicator.}
\end{deluxetable}

\begin{deluxetable}{lccccccc}
\tablewidth{0pt}
\tablecaption{X-ray Spectral Analysis
\label{xspec}}
\tablehead{
\colhead{Source} & \colhead{Net Counts} & \colhead{$N_H$} & \colhead{k$T_1$}  & \colhead{k$T_2$ }& \colhead{$\chi^2_{red}$}& \colhead{ Flux}\\
\colhead{ } &\colhead{ }      &\colhead{$10^{21}\; \rm{cm}^{-2}$ } &\colhead{ keV}      & \colhead{keV} & \colhead{ }& \colhead{$\rm{erg\;s^{-1}\;cm^{-2}}$}  \\}
\startdata
CR Cha&1996&0.8&0.84&2.97&0.95&1.4$\times10^{-12}$\\
CHXR 30A&324&12.2&&2.14&1.01&4.6$\times10^{-13}$\\
SY Cha&283&0.8&0.34&2.94&0.81&2.4$\times10^{-13}$\\
SZ Cha&692&0.8&0.85&4.53&0.74&3.7$\times10^{-13}$\\
TW Cha&313&0.8&0.41&1.95&1.08&1.8$\times10^{-13}$\\
CVSO 38&266&1.1 &0.98&6.98&1.42&4.1$\times10^{-13}$\\
J05244498+0159465&285&1.1&0.32&1.59&2.06&2.5$\times10^{-13}$\\
RX J0526+0143&369&1.1&0.31&1.48&2.29&2.6$\times10^{-13}$\\
\enddata
\end{deluxetable}




\begin{thebibliography}

\bibitem[Abgrall et al.(1997)]{abgrall97} Abgrall, H., Roueff, 
E., Liu, X., \& Shemansky, D.~E.\ 1997, \apj, 481, 557

\bibitem[Alcala et 
al.(1996)]{alcala96} Alcala, J.~M., et al.\ 1996, \aaps, 119, 7

\bibitem[Alcal{\'a} et al.(2008)]{alcala08} Alcal{\'a}, J.~M., 
et al.\ 2008, \apj, 676, 427

\bibitem[Alexander et al.(2005)]{alexander05} Alexander, R.~D., 
Clarke, C.~J., \& Pringle, J.~E.\ 2005, \mnras, 358, 283

\bibitem[Alexander et al.(2006)]{alexander06} Alexander, R.~D., 
Clarke, C.~J., \& Pringle, J.~E.\ 2006, \mnras, 369, 229

\bibitem[Alexander 
\& Armitage(2007)]{alexander07} Alexander, R.~D., \& Armitage, P.~J.\ 2007, \mnras, 375, 500

\bibitem[Ardila et al.(2002)]{ardila2002} Ardila, D.~R., Basri, 
G., Walter, F.~M., Valenti, J.~A., 
\& Johns-Krull, C.~M.\ 2002, \apj, 566, 1100 

\bibitem[Arnaud(1996)]{arnaud96} Arnaud, K.~A.\ 1996, 
Astronomical Data Analysis Software and Systems V, 101, 17

\bibitem[Bergin et al.(2004)]{bergin04} Bergin, E., et al.\ 
2004, \apjl, 614, L133

\bibitem[Beristain et al.(2001)]{beristain2001} Beristain, G., 
Edwards, S., \& Kwan, J.\ 2001, \apj, 551, 1037

\bibitem[Brice{\~n}o et al.(2005)]{briceno05} Brice{\~n}o, C., 
Calvet, N., Hern{\'a}ndez, J., Vivas, A.~K., Hartmann, L., Downes, J.~J., 
\& Berlind, P.\ 2005, \aj, 129, 907

\bibitem[Brice{\~n}o et al.(2007)]{briceno07} Brice{\~n}o, C., 
Hartmann, L., Hern{\'a}ndez, J., Calvet, N., Vivas, A.~K., Furesz, G., 
\& Szentgyorgyi, A.\ 2007, \apj, 661, 1119

\bibitem[Brice{\~n}o et al.(2010)]{briceno10} Brice{\~n}o, C., et al. \ 2010 (in prep.)

\bibitem[Brickhouse et al.(2010)]{brickhouse10} Brickhouse, N.~S., 
Cranmer, S.~R., Dupree, A.~K., Luna, G.~J.~M., 
\& Wolk, S.\ 2010, \apj, 710, 1835

\bibitem[Briggs 
\& Pye(2003)]{briggs03} Briggs, K.~R., \& Pye, J.~P.\ 2003, \mnras, 345, 714

\bibitem[Byrne \& Doyle(1989)]{byrne1989} Byrne, P.~B., \& Doyle, J.~G.\ 1989, \aap, 208, 159

\bibitem[Calvet 
\& Gullbring(1998)]{calvet98} Calvet, N., \& Gullbring, E.\ 1998, \apj, 509, 802

\bibitem[Calvet et al.(2002)]{calvet02} Calvet, N., D'Alessio, P., Hartmann, L., Wilner, D., Walsh, A., \& Sitko, M.\ 2002, \apj, 568, 1008

\bibitem[Calvet et al.(2004)]{calvet04} Calvet, N., Muzerolle, 
J., Brice{\~n}o, C., Hern{\'a}ndez, J., Hartmann, L., Saucedo, J.~L., 
\& Gordon, K.~D.\ 2004, \aj, 128, 1294 

\bibitem[Calvet et al.(2005)]{calvet05} Calvet, N., Brice{\~n}o, C., Hern{\'a}ndez, J., Hoyer, S., Hartmann, L., Sicilia-Aguilar, A., Megeath, S.~T., \& D'Alessio, P.\ 2005, \aj, 129, 935


\bibitem[Carpenter et al.(2008)]{carpenter08} Carpenter, J.~M., et 
al.\ 2008, \apjs, 179, 423

\bibitem[Chen et al.(2005)]{chen05} Chen, C.~H., et al.\ 2005, 
\apj, 634, 1372

\bibitem[Clarke et al.(2001)]{clarke01} Clarke, C.~J., Gendrin, 
A., \& Sotomayor, M.\ 2001, \mnras, 328, 485

\bibitem[Currie(2008)]{currie08} Currie, T.~M.\ 2008, 
Ph.D.~Thesis

\bibitem[D'Alessio et al.(2006)]{dalessio06} D'Alessio, P., 
Calvet, N., Hartmann, L., Franco-Hern{\'a}ndez, R., 
\& Serv{\'{\i}}n, H.\ 2006, \apj, 638, 314

\bibitem[Dickey 
\& Lockman(1990)]{dickey90} Dickey, J.~M., \& Lockman, F.~J.\ 1990, \araa, 28, 215

\bibitem[Dullemond 
\& Dominik(2005)]{dullemond05} Dullemond, C.~P., \& Dominik, C.\ 2005, \aap, 434, 971

\bibitem[Edwards et al.(2006)]{edwards2006} Edwards, S., Fischer, 
W., Hillenbrand, L., \& Kwan, J.\ 2006, \apj, 646, 319 

\bibitem[Espaillat et al.(2007)]{espaillat07} Espaillat, C., 
Calvet, N., D'Alessio, P., Hern{\'a}ndez, J., Qi, C., Hartmann, L., Furlan, 
E., \& Watson, D.~M.\ 2007, \apjl, 670, L135

\bibitem[Espaillat et al.(2008)]{espaillat08} Espaillat, C., et 
al.\ 2008, \apjl, 689, L145

\bibitem[Espaillat(2009)]{espaillat09} Espaillat, C.\ 2009, 
Ph.D.~Thesis

\bibitem[Espaillat et al.(2010)]{espaillat10} Espaillat, C., et 
al.\ 2010, \apj, 717, 441

\bibitem[Fedele et 
al.(2010)]{fedele10} Fedele, D., van den Ancker, M.~E., Henning, T., Jayawardhana, R., \& Oliveira, J.~M.\ 2010, \aap, 510, A72

\bibitem[Feigelson et al.(1993)]{feigelson93} Feigelson, E.~D., 
Casanova, S., Montmerle, T., \& Guibert, J.\ 1993, \apj, 416, 623\

\bibitem[Feigelson et al.(2002)]{feigelson02} Feigelson, E.~D., 
Broos, P., Gaffney, J.~A., III, Garmire, G., Hillenbrand, L.~A., Pravdo, 
S.~H., Townsley, L., \& Tsuboi, Y.\ 2002, \apj, 574, 258\

\bibitem[Feigelson et al.(2005)]{feigelson05} Feigelson, E.~D., et 
al.\ 2005, \apjs, 160, 379

\bibitem[France et al.(2010)]{france10} France, K., Linsky, 
J.~L., Brown, A., Froning, C.~S., \& B{\'e}land, S.\ 2010, \apj, 715, 596

\bibitem[Fruscione et al.(2006)]{fruscione06} Fruscione, A., et 
al.\ 2006, \procspie, 6270

\bibitem[Furlan et al.(2009)]{furlan09} Furlan, E., et al.\ 
2009, \apj, 703, 1964

\bibitem[Gauvin 
\& Strom(1992)]{gauvin92} Gauvin, L.~S., \& Strom, K.~M.\ 1992, \apj, 385, 217 

\bibitem[Getman et al.(2008)]{getman08} Getman, K.~V., 
Feigelson, E.~D., Broos, P.~S., Micela, G., 
\& Garmire, G.~P.\ 2008, \apj, 688, 418

\bibitem[Gorti 
\& Hollenbach(2009a)]{gorti09} Gorti, U., \& Hollenbach, D.\ 2009, \apj, 690, 1539

\bibitem[Gorti et al.(2009b)]{gorti09b} Gorti, U., Dullemond, 
C.~P., \& Hollenbach, D.\ 2009, \apj, 705, 1237

\bibitem[G{\"u}del et 
al.(2007)]{gudel07} G{\"u}del, M., et al.\ 2007, \aap, 468, 353

\bibitem[G{\"u}nther et 
al.(2007)]{gunther07} G{\"u}nther, H.~M., Schmitt, J.~H.~M.~M., Robrade, J., \& Liefke, C.\ 2007, \aap, 466, 1111

\bibitem[Hartmann et al.(1979)]{hartmann1979} Hartmann, L., 
Schmidtke, P.~C., Davis, R., Dupree, A.~K., Raymond, J., 
\& Wing, R.~F.\ 1979, \apjl, 233, L69

\bibitem[Hartmann et al.(1998)]{hartmann98} Hartmann, L., Calvet, 
N., Gullbring, E., \& D'Alessio, P.\ 1998, \apj, 495, 385

\bibitem[Hartmann(2003)]{hartmann03} Hartmann, L.\ 2003, \apj, 
585, 398

\bibitem[Hartmann et al.(2005)]{hartmann05} Hartmann, L., Megeath, 
S.~T., Allen, L., Luhman, K., Calvet, N., D'Alessio, P., Franco-Hernandez, 
R., \& Fazio, G.\ 2005, \apj, 629, 881


\bibitem[Herczeg et al.(2002)]{herczeg02} Herczeg, G.~J., Linsky, 
J.~L., Valenti, J.~A., Johns-Krull, C.~M., 
\& Wood, B.~E.\ 2002, \apj, 572, 310

\bibitem[Herczeg et al.(2004)]{herczeg04} Herczeg, G.~J., Wood, 
B.~E., Linsky, J.~L., Valenti, J.~A., 
\& Johns-Krull, C.~M.\ 2004, \apj, 607, 369

\bibitem[Herczeg et al.(2006)]{herczeg06} Herczeg, G.~J., Linsky, 
J.~L., Walter, F.~M., Gahm, G.~F., 
\& Johns-Krull, C.~M.\ 2006, \apjs, 165, 256 

\bibitem[Hern{\'a}ndez et al.(2006)]{hernandez06} Hern{\'a}ndez, 
J., Brice{\~n}o, C., Calvet, N., Hartmann, L., Muzerolle, J., 
\& Quintero, A.\ 2006, \apj, 652, 472

\bibitem[Hern{\'a}ndez et al.(2007a)]{hernandez07} Hern{\'a}ndez, 
J., et al.\ 2007, \apj, 671, 1784

\bibitem[Hern{\'a}ndez et al.(2007b)]{hernandez07b} Hern{\'a}ndez, 
J., et al.\ 2007, \apj, 662, 1067

\bibitem[Hern{\'a}ndez et al.(2009)]{hernandez09} Hern{\'a}ndez, 
J., Calvet, N., Hartmann, L., Muzerolle, J., Gutermuth, R., 
\& Stauffer, J.\ 2009, \apj, 707, 705 

\bibitem[Hern{\'a}ndez et al.(2010)]{hernandez10} Hern{\'a}ndez, 
J., Morales-Calderon, M., Calvet, N., Hartmann, L., Muzerolle, J., 
Gutermuth, R., Luhman, K.~L., \& Stauffer, J.\ 2010, \apj, 722, 1226

\bibitem[Hollenbach et al.(1994)]{hollenbach94} Hollenbach, D., 
Johnstone, D., Lizano, S., \& Shu, F.\ 1994, \apj, 428, 654

\bibitem[Huenemoerder et al.(1994)]{huenemoerder94} Huenemoerder, 
D.~P., Lawson, W.~A., \& Feigelson, E.~D.\ 1994, \mnras, 271, 967

\bibitem[Ingleby et al.(2009)]{ingleby09} Ingleby, L., et al.\ 
2009, \apjl, 703, L137

\bibitem[Ingleby et a.(2011)]{ingleby11} Ingleby, L., et al. (in prep.)

\bibitem[Johns-Krull et al.(2000)]{johnskrull00} Johns-Krull, C.~M., 
Valenti, J.~A., \& Linsky, J.~L.\ 2000, \apj, 539, 815

\bibitem[Kalas et al.(2006)]{kalas06} Kalas, P., Graham, J.~R., 
Clampin, M.~C., \& Fitzgerald, M.~P.\ 2006, \apjl, 637, L57

\bibitem[Kalberla et al.(2005)]{kalberla05} Kalberla, P.~M.~W., Burton, W.~B., Hartmann, D., Arnal, E.~M., Bajaja, E., Morras, R., Poppel, W.~G.~L.\ 2005, \aap, 440, 775

\bibitem[Kastner et al.(2003)]{kastner03} Kastner, J.~H., 
Crigger, L., Rich, M., \& Weintraub, D.~A.\ 2003, \apj, 585, 878

\bibitem[Kenyon 
\& Hartmann(1995)]{kenyon95} Kenyon, S.~J., \& Hartmann, L.\ 1995, \apjs, 101, 117

\bibitem[Kim et al.(2009)]{kim09} Kim, K.~H., et al.\ 2009, 
\apj, 700, 1017

\bibitem[Lopez-Santiago et al.(2010)]{lopez10} Lopez-Santiago, 
J., Albacete Colombo, J.~F., \& Lopez-Garcia, M.~A.\ 2010, arXiv:1009.3392

\bibitem[Luhman(2004)]{luhman04} Luhman, K.~L.\ 2004, \apj, 602, 
816

\bibitem[Luhman(2007)]{luhman07} Luhman, K.~L.\ 2007, \apjs, 
173, 104

\bibitem[Luhman et al.(2008)]{luhman08} Luhman, K.~L., et al.\ 
2008, \apj, 675, 1375

\bibitem[Luhman et al.(2010)]{luhman10} Luhman, K.~L., Allen, 
P.~R., Espaillat, C., Hartmann, L., \& Calvet, N.\ 2010, \apjs, 186, 111\

\bibitem[Low et al.(2005)]{low05} Low, F.~J., Smith, P.~S., 
Werner, M., Chen, C., Krause, V., Jura, M., 
\& Hines, D.~C.\ 2005, \apj, 631, 1170

\bibitem[Manoj(2010)]{manoj10} Manoj, P.\ 2010, arXiv:1003.5933

\bibitem[Marshall et al.(2008)]{marshall08} Marshall, J.~L., et 
al.\ 2008, \procspie, 7014

\bibitem[Mathis(1990)]{mathis90} Mathis, J.~S.\ 1990, \araa, 28, 37

\bibitem[Mercer et al.(2009)]{mercer09} Mercer, E.~P., Miller, 
J.~M., Calvet, N., Hartmann, L., Hernandez, J., Sicilia-Aguilar, A., 
\& Gutermuth, R.\ 2009, \aj, 138, 7

\bibitem[Muzerolle et al.(2010)]{muzerolle10} Muzerolle, J., Allen, 
L.~E., Megeath, S.~T., Hern{\'a}ndez, J., 
\& Gutermuth, R.~A.\ 2010, \apj, 708, 1107

\bibitem[Najita et al.(2010)]{najita10} Najita, J.~R., Carr, 
J.~S., Strom, S.~E., Watson, D.~M., Pascucci, I., Hollenbach, D., Gorti, 
U., \& Keller, L.\ 2010, \apj, 712, 274

\bibitem[Natta et 
al.(2004)]{natta04} Natta, A., Testi, L., Muzerolle, J., Randich, S., Comer{\'o}n, F., \& Persi, P.\ 2004, \aap, 424, 603

\bibitem[Nguyen et al.(2009)]{nguyen09} Nguyen, D.~C., 
Jayawardhana, R., van Kerkwijk, M.~H., Brandeker, A., Scholz, A., 
\& Damjanov, I.\ 2009, \apj, 695, 1648

\bibitem[Owen et al.(2010)]{owen10} Owen, J.~E., Ercolano, B., 
Clarke, C.~J., \& Alexander, R.~D.\ 2010, \mnras, 401, 1415

\bibitem[Pascucci et al.(2006)]{pascucci06} Pascucci, I., et al.\ 
2006, \apj, 651, 1177

\bibitem[Preibisch 
\& Feigelson(2005)]{preibisch05} Preibisch, T., \& Feigelson, E.~D.\ 2005, \apjs, 160, 390

\bibitem[Ribas et al.(2005)]{ribas05} Ribas, I., Guinan, E.~F., 
G{\"u}del, M., \& Audard, M.\ 2005, \apj, 622, 680

\bibitem[Robitaille et al.(2007)]{robitaille07} Robitaille, T.~P., 
Whitney, B.~A., Indebetouw, R., \& Wood, K.\ 2007, \apjs, 169, 328

\bibitem[Robrade 
\& Schmitt(2006)]{robrade06} Robrade, J., \& Schmitt, J.~H.~M.~M.\ 2006, \aap, 449, 737

\bibitem[Sacco et 
al.(2008)]{sacco08} Sacco, G.~G., Argiroffi, C., Orlando, S., Maggio, A., Peres, G., \& Reale, F.\ 2008, \aap, 491, L17

\bibitem[Sargent et al.(2006)]{sargent06} Sargent, B., et al.\ 
2006, \apj, 645, 395

\bibitem[Sicilia-Aguilar et al.(2006)]{sicilia06} 
Sicilia-Aguilar, A., Hartmann, L.~W., F{\"u}r{\'e}sz, G., Henning, T., 
Dullemond, C., \& Brandner, W.\ 2006, \aj, 132, 2135

\bibitem[Sicilia-Aguilar et al.(2010)]{sicilia10} 
Sicilia-Aguilar, A., Henning, T., \& Hartmann, L.~W.\ 2010, \apj, 710, 597

\bibitem[Siess et al.(2000)]{siess00} Siess, L., Dufour, E., \& Forestini, M.\ 2000, \aap, 358, 593

\bibitem[Song et al.(2000)]{song00} Song, I., Caillault, 
J.-P., Barrado y Navascu{\'e}s, D., Stauffer, J.~R., 
\& Randich, S.\ 2000, \apjl, 533, L41

\bibitem[Spezzi et al.(2008)]{spezzi08} Spezzi, L., et al.\ 
2008, \apj, 680, 1295

\bibitem[Strom et al.(1989)]{strom89} Strom, K.~M., Strom, 
S.~E., Edwards, S., Cabrit, S., \& Skrutskie, M.~F.\ 1989, \aj, 97, 1451

\bibitem[Skrutskie et al.(2006)]{skrutskie06} Skrutskie, M.~F., et 
al.\ 2006, \aj, 131, 1163

\bibitem[Tody(1993)]{tody93} Tody, D.\ 1993, Astronomical Data 
Analysis Software and Systems II, 52, 173

\bibitem[Valenti et al.(2000)]{valenti00} Valenti, J.~A., 
Johns-Krull, C.~M., \& Linsky, J.~L.\ 2000, \apjs, 129, 399

\bibitem[Voges et 
al.(1999)]{voges99} Voges, W., et al.\ 1999, \aap, 349, 389 

\bibitem[Voges et al.(2000)]{voges00} Voges, W., et al.\ 2000, 
VizieR Online Data Catalog, 9029, 0 

\bibitem[Webb et al.(1999)]{webb99} Webb, R.~A., Zuckerman, 
B., Platais, I., Patience, J., White, R.~J., Schwartz, M.~J., 
\& McCarthy, C.\ 1999, \apjl, 512, L63

\bibitem[Whittet et al.(2004)]{whittet04} Whittet, D.~C.~B., 
Shenoy, S.~S., Clayton, G.~C., \& Gordon, K.~D.\ 2004, \apj, 602, 291

\bibitem[Wichmann et 
al.(2003)]{wichmann03} Wichmann, R., Schmitt, J.~H.~M.~M., \& Hubrig, S.\ 2003, \aap, 399, 983



\end{thebibliography}
\end{document}